\documentclass[useAMS,usenatbib]{mn2e}
\usepackage{epsfig}
\usepackage{times}

\newcommand{\B}[1]{\mathbf{#1}}

\newcommand{\Bx}{\B{x}} \newcommand{\By}{\B{y}}
 
\newcommand{\Bv}{\B{v}}

\newcommand{\cf}{\bar{f}}
\newcommand{\cv}{\bar{v}}

\newcommand{\Eq}[1]{Eq.~(\ref{#1})} 

\def\equ#1{eq.~(\ref{eq:#1})}

\def\se#1{\S\ref{sec:#1}}
\def\Fig#1{Fig.~\ref{fig:#1}}
\def\Figu#1{Figure~\ref{fig:#1}}

\def\\{\hfill\break}

\def\prop{\propto}
\def\ifm#1{\relax\ifmmode#1\else$\mathsurround=0pt #1$\fi}
\def\kms{\ifmmode\,{\rm km}\,{\rm s}^{-1}\else km$\,$s$^{-1}$\fi}
\def\hmpc{\,\ifm{h^{-1}}{\rm Mpc}}
\def\mpc{\,{\rm Mpc}}

\def\kpc{\,{\rm kpc}}

\def\msun{M_{\odot}}
\def\ltsima{$\; \buildrel < \over \sim \;$}
\def\lsim{\lower.5ex\hbox{\ltsima}}
\def\gtsima{$\; \buildrel > \over \sim \;$}
\def\gsim{\lower.5ex\hbox{\gtsima}}
\def\gtrsim{\lower.5ex\hbox{\gtsima}}
\def\rarrow{\rightarrow}

\def\Rvir{R_{\rm vir}}
\def\Mvir{M_{\rm vir}}
\def\Nvir{M_{\rm vir}}

\def\fvir{f_{\rm vir}}
\def\f02{f_{20\%}}
\def\fdel{f_{\rm del}}
\def\vdel{v_{\rm del}}
\def\Vdel{V_{\rm del}}
\def\fvor{f_{\rm vor}}
\def\vvor{v_{\rm vor}}
\def\Vvor{V_{\rm vor}}
\def\ftrue{f_{\rm true}}
\def\vtrue{v_{\rm true}}

\def\sigv{\sigma_{\rm v}}
\def\sigc{\sigma_{\rm c}} 

\def\pmb#1{\setbox0=\hbox{#1}
 \kern-.025em\copy0\kern-\wd0
 \kern.05em\copy0\kern-\wd0
 \kern-.025em\raise.0433em\box0}
\def\boldvf{\pmb{$v(f)$}}

\def\boldvdel{\pmb{$\vdel(f)$}}

\title[Phase-Space Structure of Dark-Matter Haloes]
{Phase-Space Structure of Dark-Matter Haloes:\\
Scale-Invariant PDF Driven by Substructure}

\author[I. Arad, A. Dekel, A. Klypin]
{ Itai Arad$^1$, Avishai Dekel$^2$, Anatoly Klypin$^3$\\
  $^1$Institute of Astronomy, Madingley Road, Cambridge CB3 OHA, UK,
  arad@ast.cam.ac.uk\\
  $^2$Racah Institute of Physics, The Hebrew University,
  Jerusalem 91904, Israel, dekel@phys.huji.ac.il\\
  $^3$Astronomy Department, New Mexico State University, Box 30001,
  Dept.  4500, Las Cruces, NM 88003, USA, aklypin@nmsu.edu}

\begin{document}

\pagerange{\pageref{firstpage}--\pageref{lastpage}} \pubyear{2003}
\maketitle
\label{firstpage}

\begin{abstract}
  
  We present a method for computing the 6-dimensional coarse-grained
  phase-space density $f(\Bx,\Bv)$ in an N-body system, and derive its
  distribution function $v(f)$.  The method is based on Delaunay
  tessellation, where $v(f)$ is obtained with an effective fixed
  smoothing window over a wide $f$ range.  The errors are estimated,
  and $v(f)$ is found to be insensitive to the sampling resolution or
  the simulation technique.  We find that in gravitationally relaxed
  haloes built by hierarchical clustering, $v(f)$ is well approximated
  by a robust power law, $v(f) \propto f^{-2.5 \pm 0.05}$, over more
  than 4 decades in $f$, from its virial level to the numerical
  resolution limit. This is tested to be valid in the $\Lambda$CDM
  cosmology for haloes with masses $10^9-10^{15}\msun$, indicating
  insensitivity to the slope of the initial fluctuation power
  spectrum.  By mapping the phase-space density in position space, we
  find that the high-$f$ end of $v(f)$ is dominated by the ``cold''
  subhaloes rather than the parent-halo central region and its global
  spherical profile.  The value of $f$ in subhaloes near the virial
  radius is typically $>100$ times higher than its value at the halo
  centre, and it decreases gradually from outside in toward its value
  at the halo centre.  This seems to reflect phase mixing due to
  mergers and tidal effects involving puffing up and heating.  The
  phase-space density can thus provide a sensitive tool for studying
  the evolution of subhaloes during the hierarchical buildup of
  haloes.  It remains to be understood why the evolved substructure
  adds up to the actual universal power law of $v(f) \propto
  f^{-2.5}$.  It seems that this behaviour results from the
  hierarchical clustering process and is not a general result of
  violent relaxation.

\end{abstract}


\section{INTRODUCTION}
\label{sec:intro}

The standard paradigm assumes that dark-matter haloes are the basic
entities in which luminous galaxies form and live.  The haloes
dominate the gravitational potential over a wide range of radii and
they have a crucial role in determining the galaxy properties. While
many of the systematic features of halo structure and kinematics have
been revealed by $N$-body simulations, the origin of these features is
still an open theoretical puzzle, despite the fact that they are due
to simple Newtonian gravity.

The halo \emph{density profile} $\rho(r)$ is an example for such a
puzzle.  It is found in the simulations to have a robust non-power-law
shape (which we refer to in general as ``NFW"), with a log slope $-3$
near the virial radius, flattening gradually toward $-1$ at about 1\%
of the virial radius, and perhaps flattening further at smaller radii.
\citep[e.g.][and references therein]{ref:Nav97,ref:Mo98,ref:Kly01,
  ref:Pow03,ref:Hay04}.  This density profile is insensitive (or at
most weakly sensitive) to parameters of the cosmological model and the
initial fluctuation power spectrum
\citep[e.g.,][]{ref:Col96,ref:Nav97,ref:Sub00, ref:Ric03, ref:Col03,
  ref:Nav04} indicating that its origin is due to a robust relaxation
process rather than specific initial conditions.  In particular,
violent relaxation \citep{ref:Lyn67} may be involved in shaping up the
density profile.  In addition, secondary infall may be important in
the outer regions \citep{ref:Gun72, ref:Dek81, ref:Fil84, ref:Hof85,
  ref:Whi92} while some argue that resonances may have a role in the
inner regions (\citealt{ref:Wei02}; but see \citealt{ref:Val03}).
Nevertheless, there is no clear understanding for why the haloes
actually pick up the particular density profile they have.

The properties of the \emph{velocity dispersion} tensor is another
theoretical puzzle. The haloes tend to be rather round, with a
velocity dispersion profile that is slightly rising at small radii and
slightly falling at large radii but is rather flat overall
\citep{ref:Hus99a, ref:Hus99b}. The profile of the anisotropy
parameter $\beta(r)$, which is a measure of radial versus tangential
velocities, indicates near isotropy at small radii, which gradually
develops into more radial orbits at large radii \citep{ref:Col96,
  ref:Hus99a, ref:Hus99b, ref:Col00a}.  For a spherical system in
equilibrium, the $\sigma(r)$ and $\beta(r)$ are related to $\rho(r)$
via the Jeans equation, but it is not at all clear why the haloes in
the simulations choose the characteristic $\sigma(r)$ or $\beta(r)$
that they actually do.

An interesting attempt to address the origin of the halo profile has
been made by \citet{ref:Tay01}, who measured a poor-man phase-space
density profile by $f_{\rm TN}(r) =\rho(r)/\sigma(r)^3$, and found
that it displays an approximate power-law behaviour, $f_{\rm TN}
\propto r^{-1.87}$, over more than two decades in $r$.  Using the
spheri-symmetric Jeans equation, they showed that this power law
permits a whole family of density profiles, whose limiting case is a
profile similar to NFW, which asymptotically approaches a slope
$-0.75$ as $r \rarrow 0$. The general power-law shape of $f_{\rm
  TN}(r)$ is confirmed in the simulated haloes described below. This
scale-free behaviour of $f_{\rm TN}(r)$ is intriguing, and it
motivates further studies of halo structure by means of phase-space
density.

Simulations of the $\Lambda$CDM cosmology also reveal a roughly
self-similar \emph{hierarchical clustering} process, where smaller
building blocks accrete and merge into bigger haloes.  At every
moment, every halo contains a substructure of subhaloes on top of a
smooth halo component that has been tidally stripped from an earlier
generation of substructure.  Some of the important dynamical processes
involved in this hierarchical halo buildup are understood
qualitatively. These include, for example, the dynamical friction
which governs the decay of the satellite orbits, the tidal stripping
of subhaloes due to the host halo potential, and the mergers and flyby
interactions of the subhaloes among themselves.  However, a complete
theoretical understanding of how these processes work in detail, and
how they combine to produce the halo structure and kinematics, is
lacking.

Attempts have been made to explain an inner density cusp using toy
models of dynamical stripping and tidal effects during the halo
buildup by mergers \citep{ref:Syr98,ref:Nus99,ref:ddh03,ref:Dek03}.
However, a similar halo density profile seems to be produced also in
some of the simulations where substructure has been artificially
suppressed \citep{ref:Mo99b, ref:Hus99a, ref:Avila01, ref:Bul02,
  ref:Alvarez02}, indicating that the process responsible for the
origin of this density profile might be a somewhat more general
feature of gravity and not unique to the merger scenario.

The issue of halo substructure has become timely both because of its
relevance to observations and its implications on other major issues
in galaxy formation.  Tidal tails and streams associated with dwarf
satellite galaxies have been observed in the haloes of the Milky Way
and M31 \citep{ref:Iba94, ref:Iba01a, ref:Iba01b}, and they start to
allow detailed modelling of the halo history through the satellite
orbits. Gravitational-lensing observations provide preliminary
indications for the presence of substructure in haloes at the high
level predicted by the dissipationless $\Lambda$CDM scenario
\cite[e.g.][]{ref:Dal02} In contrast, the observed number density of
dwarf galaxies seems to be significantly lower, thus posing a
``missing dwarf problem'' \cite{ref:Kly99b, ref:Mo99a}. Also, the
``angular-momentum problem'' of disk galaxies
\citep[e.g.][]{ref:Nav00,ref:Bul01a} is probably associated with the
evolution of substructure in haloes \citep{ref:Mal02, ref:mds02}.
While the dwarf and angular-momentum problems necessarily involve
baryonic processes, understanding the gravitational evolution of
substructure is clearly a key for solving them.

In order to better understand the origin the various aspects of halo
structure and buildup mentioned above, we make here a first attempt at
addressing directly and in some detail the \emph{phase-space}
structure of dark-matter haloes.  The fundamental quantity in the
dynamical evolution of gravitating systems is the full,
six-dimensional, coarse-grained, phase-space density $f(\Bx,\Bv)$,
which intimately relates to the underlying Vlasov equation and lies
behind any relaxation process that may give rise to the virialized
halo structure \citep[][ chapter 4]{ref:Bin87}.

Ideally, one would have liked to compute it free of assumptions
concerning spherical symmetry, isotropy, or any kind of equilibrium.
However, computing densities in a six-dimensional space is a
non-trivial challenge which requires simulations of a very broad
dynamical range.  The state-of-the-art N-body simulations, with more
than million particles per halo, allow an attempt of this sort for the
first time.  We describe below a successful algorithm for measuring
$f(\Bx,\Bv)$, and study its relevant properties including the
associated systematic and random uncertainties.  We then apply this
algorithm to simulated virialized haloes in the $\Lambda$CDM
cosmology.

We report in this paper two surprising new results. First, we discover
that the volume distribution function of the phase-space density,
$v(f)$, displays a universal scale-invariant \emph{power-law} shape,
valid in all virialized haloes that form by hierarchical clustering.
Second, we realise that this power law is not directly related to the
overall density profile, but is rather driven by the halo
\emph{substructure}.  This implies that the phase-space density
provides a useful tool for studying the hierarchical buildup of
dark-matter haloes and the evolution of substructure in them.

In \se{vf} we introduce $f(\Bx,\Bv)$ and $v(f)$.  In \se{computing} we
describe the method of computing $f$ and $v(f)$ from an $N$-body halo,
and summarise its properties and the associated errors, which are
addressed in more detail in Appendix~\ref{sec:errors}.  In
\se{universal} we present the universal power-law shape of $v(f)$
based on several different simulations, and demonstrate its robustness
to the mass scale and simulation technique.  In \se{substructures} we
display maps of phase-space density and show that the high-$f$
contributions to $v(f)$ come from substructures within the parent
halo.  In \se{conc} we summarise and discuss our results and future
work.

\section{THE PDF OF PHASE-SPACE DENSITY: \boldvf}
\label{sec:vf}

This is a more technical and elaborate introductory section, aimed at
introducing the concepts and nomenclature relevant for the analysis of
this paper.

\subsection{Definitions}

The state of a collisionless system is completely determined by the
fine-grained phase-space density function $f(\Bx,\Bv,t)$, which
measures the mass contained in an infinitesimal phase-space patch of
volume $d\Bx d\Bv$, located at $(\Bx,\Bv)$. The evolution of
$f(\Bx,\Bv,t)$ is governed by the Vlasov equation,
\begin{equation}
  \label{eq:vlasov}
  \partial_t f + \Bv \cdot \nabla_{\Bx} f 
    - \nabla_{\Bx}\Phi \cdot \nabla_\Bv f = 0 \ , 
\end{equation}
with $\Phi(\Bx)$ the gravitational potential, related
self-consistently to $f(\Bx,\Bv)$ by the Poisson's integral
\begin{equation}
  \Phi(\Bx) = -G\int \!\! d\Bx' d\Bv \, 
                \frac{f(\Bx',\Bv)}{|\Bx-\Bx'|} \ .
\end{equation}

It is therefore sensible to assume that a true understanding of the
nature of self-gravitating collisionless systems must involve
$f(\Bx,\Bv)$ as a primary ingredient. However, being a function of six
variables, $f(\Bx,\Bv)$ is hard to deal with computationally.  It
turns out that there is a much simpler function, $v(f)$, which is
intimately related to $f(\Bx,\Bv)$, but is much easier to handle both
analytically and numerically.

Before defining $v(f)$, it is worth recalling that the Vlasov equation
preserves the phase-space density along particle trajectories: It
implies \equ{vlasov} implies that
\begin{equation}
  \frac{d}{dt}f[\Bx(t), \Bv(t)] = 0 \ ,
\end{equation} 
where $[\Bx(t), \Bv(t)]$ is the trajectory of a phase-space element.
This conservation of phase-space density along trajectories can also
be viewed as a conservation of phase-space volume.  In order to
illustrate what this means, consider any smooth function $C(x)$, and
define the integral
\begin{equation}
  \label{eq:int-I}
  I(t) \equiv \int\!\!d\Bx d\Bv \, C[f(\Bx,\Bv,t)] \ .
\end{equation}
Using the Vlasov equation we find that
\begin{eqnarray}
  \frac{dI}{dt} &=& \int\!\!d\Bx d\Bv \, C'(f)\partial_t f \\
     &=& -\int\!\!d\Bx d\Bv \, C'(f)\big[\Bv \cdot \nabla_{\Bx} f 
    - \nabla_{\Bx}\Phi \cdot \nabla_\Bv f\big] \\
  &=& -\int\!\!d\Bx d\Bv \, \big[\Bv \cdot \nabla_{\Bx} C(f) 
    - \nabla_{\Bx}\Phi \cdot \nabla_\Bv C(f)\big] = 0 \ .
\label{eq:dC}
\end{eqnarray}
The last equality follows by integrating the first term over $\Bx$ and
the second term over $\Bv$, and assuming that $f\to 0$ as $\Bx, \Bv
\to \infty$. This implies that $I$ is conserved under the dynamics.

Now, if $C(f)$ is the Dirac delta 
function\footnote{Strictly speaking, the Dirac delta function is not a
smooth function, but it can be approximated by a series of smooth
functions, each obeying \Eq{eq:dC}} 
$\delta(f-f_0)$, then the integral $I$ becomes an $f$-dependent
function, which we refer to as the \emph{PDF} $v(f)$:
\begin{equation}
  \label{def:vf}
  v(f=f_0) \equiv \int \!\! d\Bx d\Bv\, \delta\big[ f(\Bx, \Bv, t) - f_0\big]
  \ .
\end{equation}
From this definition it is clear that $v(f)df$ is the volume of
phase-space occupied by phase-space elements whose density lies in the
range $(f,f+df)$. The conservation of $v(f)$ implied by \Eq{eq:dC}
means that this volume is conserved under the dynamics.  The
fine-grained $v(f)$ [i.e., the $v(f)$ which is calculated using the
fine-grained $f(\Bx,\Bv)$] should therefore be viewed as a signature
of the system, which remains the same throughout its evolution.

\subsection{Coarse-Grained Density and Mixing}

One might have thought that the conservation of $v(f)$ poses severe
constrains on the evolution of $f(\Bx,\Bv,t)$, but this is not the
case due to the effects of \emph{mixing}.  In the course of evolution,
phase-space patches of high $f$ are stretched and spiral into regions
with low $f$. As the stretching continues, the patches become thiner
and thiner, and as a result $f(\Bx, \Bv)$ varies over increasingly
smaller scales. Very soon, one can no longer measure $f(\Bx, \Bv)$,
but instead measure an average of it over some finite volume. This
average is often referred to as the ``coarse-grained" phase-space
density, as opposed to the original ``fine-grained" density $f(\Bx,
\Bv)$.  We denote the coarse-grained quantities by $\cf$ and
$\cv(\cf)$, but in subsequent sections we may omit the bar and simply
refer to them as $f$ and $v(f)$.  The important point to realise is
that the coarse-grained $\cv(\cf)$ is no longer conserved. For
example, given initially a volume $V$ filled with phase-space density
$f_A$ and a similar volume filled with density $f_B$, one ends after
mixing with a volume $2V$ filled with an average density
$(f_A+f_B)/2$.

When a large fraction of the mass is added to halo, e.g. by collapse
or a major merger, rapid global fluctuations of the gravitational
potential re-distribute the energies of each phase-space element, and
lead to very strong mixing across large scales, resulting in
variations in $\cf$.  After a few global dynamical times, the
potential fluctuations fade away and $\cf$ stabilises. The
fine-grained $f$, on the other hand, continues to spiral and stretch
indefinitely, on smaller and smaller scales, which no longer affect
$\cf$.  After a while, the $\cf$ can be viewed as the ``physical"
phase-space density of the system, since the microscopic fluctuations
of $f$ can no longer be measured, and, in particular, no longer affect
the gravitational potential. It is therefore the coarse-grained $\cf$
which, according to Jeans' theorem, can be written as a function of he
invariants, the energy $E$ and the angular momentum $\B{L}$. This
relaxation process is referred to as \emph{violent relaxation}
\citep{ref:Lyn67}.

What is the coarse-grained $\cv(\cf)$ of a relaxed system, and how is
it related to the constant fine-grained $v(f)$ of that system?  One
may identify the fine-grained $v(f)$ with the coarse-grained
$\cv(\cf)$ at an early time when the initial $f$ is smooth enough not
to vary over scales smaller than the averaging scale associated with
$\cf$. Then, the question can be rephrased in terms of the relation
between the final $\cv(\cf)$ and the early $\cv(\cf)$.  One obvious
constraint on the final $\cf$, as an average of $f$, is that their
maximum values should obey $\cf_{\rm max} \leq f_{\rm max}$.  In
particular, if the early $\cv(\cf)$ [namely $v(f)$] vanishes for
$\cf>f_{\rm max}$, then so does the final $\cv(\cf)$.

The \emph{mixing theorem} \citep{ref:Tre86} specifies 
several additional constraints on the final $\cv(\cf)$, which arise from
the fact that it has originated from a given $v(f)$.
The \emph{cumulative} distribution function is defined by
\begin{equation}
  \label{def:VF}
  V(f_0) \equiv \int\!\!d\Bx\Bv \, \theta\big[f(\Bx,\Bv)-f_0\big] 
       = \int_{f_0}^\infty \!\! v(f')\, df' \ ,
\end{equation}
where $\theta(\cdot)$ is the Heaviside step function.  As a monotonic
function, $V(f)$ can be inverted to yield $f(V)$, the \emph{reduced}
distribution function. The associated cumulative mass function $M(f)$,
which measures the mass in phase-space regions where $f(\Bx,\Bv) <
f_0$, is then defined by
\begin{eqnarray}
\label{def:MF}
M(f_0) & \equiv & \int\!\!d\Bx d\Bv \, f(\Bx,\Bv) \, 
  \theta\big[f(\Bx,\Bv)-f_0\big] \\
  & = & \int_{f_0}^\infty \!\! f' \, v(f')\, df' \ .
\end{eqnarray}
Substituting $f=f(V)$ inside $M(f)$, one obtains the function $M(V)$,
which can also be written as
\begin{equation}
  M(V) = \int_0^V \!\!f(V') \, dV' \ .
\end{equation}
If $M(V)$ and $M'(V)$ denote the functions arising from the density
functions $f(\Bx,\Bv)$ and $f'(\Bx,\Bv)$ respectively, then the mixing
theorem states that the final (coarse-grained) $\cf'$ is an average of
the initial (coarse- or fine-grained) $f$ if and only if $M'(V) \le
M(V)$ for every $V$.  This, in turn, implies an implicit constraint on
the evolution of the coarse-grained $\cv(\cf)$.  This is one of the
very few rigorous results concerning the evolution of $\cv(\cf)$, but
its strength is rather limited because it provides only an
integro-differential inequality that constrains $\cv(\cf)$.

The mixing theorem can be also stated directly in terms of $v(f)$, as
shown by \citet{ref:Mat88}. In his formulation, a necessary condition
for the final (coarse-grained) $\cv(\cf)$ evolved from the initial
$v(f)$ is that\footnote{This is the original statement in
  \citet{ref:Mat88}. However, we believe that by adding the (trivial)
  condition that $\cv(f)\ge 0$ for every $f$, it becomes also a
  sufficient condition.}
\begin{equation}
  \cv(f) = v(f) + \frac{d^2}{df^2} P(f) \ , 
\end{equation}
with $P(f)$ being some continuous function (not to be confused with a
probability distribution) such that $P(f)\le 0$ for all $f$ and
$P(f)\to 0 $ for $f\to \pm \infty$.

In the CDM scenario, the dark matter is initially cold -- with
vanishing velocity dispersion. This means a one-to-one correspondence
between the particle positions and their velocities, $\Bv=\Bv(\Bx)$,
and the initial (fine-grained) phase-space density can be written as
\begin{equation}
  f_0(\Bx,\Bv) = \rho(\Bx)\delta^3[\Bv - \Bv(\Bx)] \ . 
\end{equation}
Thus, in the beginning, the system can be described by the evolution
of its density and velocity fields, often approximated by the
Zel'dovich approximation. This singular nature of the initial
phase-space density, together with its conservation, plays a crucial
role in the formation of structures, and in particular in the
abundance of dwarf haloes, and in the formation of cusps in
dark-matter haloes \citep[e.g. ][]{ref:Col00b}.

Physically, the phase-space density is never really infinite, as it is
defined by a finite number of dark-matter particles per given
phase-space volume. Numerically, it is the finite mass-resolution of
the $N$-body simulation that prohibits the phase-space density from
being infinite. Therefore, the initial phase-space density is expected
to be extremely high in regions where $\Bv\simeq \Bv(\Bx)$ and
vanishingly small elsewhere.  Accordingly, $v(f)$ will have
contributions from a narrow range of extremely high values of $f$ and
from a narrow range near $f=0$. As the system evolves and mixing takes
place, phase-space densities from both regions mix and give rise to
intermediate values of $f$, widening the distribution $v(f)$ under the
constraints of the mixing theorem.

\subsection{A Relation Between $v(f)$ and $\rho(r)$}

In general, different systems may have the same $v(f)$.  However, if
the system is spherically symmetric and stationary, such that $f$ is a
function of the energy alone, $f(\Bx,\Bv) = f(\epsilon)$, then there
is a unique relation between $v(f)$, $f(\epsilon)$ and $\rho(r)$.  We
shall see in \se{universal} that the $v(f)$ measured in $N$-body
haloes is actually different from the $v(f)$ one would have predicted
from the halo $\rho(r)$ using the relation assuming $f(\epsilon)$.

For a spherical system the relation between $\rho(r)$ and $v(f)$ can
be derived as follows.  Assume that $f(\Bx,\Bv)=f[\epsilon(\Bx,\Bv)]$,
where $\epsilon(\Bx,\Bv)= v^2/2 + \Phi(r)$, and define the
density-of-states function
\begin{equation}
  g(\epsilon) \equiv \int\!\! d\Bx d\Bv \, 
      \delta[\epsilon(\Bx, \Bv)-\epsilon] \ . 
\end{equation}
The quantity $g(\epsilon)d\epsilon$ measures how much phase-space
volume is occupied by phase-space elements with energy in the interval
$(\epsilon,\, \epsilon+d\epsilon)$.  When the system is spherical,
$g(\epsilon)$ can be written as \citep[][Eq.~4-157b]{ref:Bin87},
\begin{equation}
  \label{def:g}
  g(\epsilon) = (4\pi)^2\int_0^{r(\epsilon)} \!\! s^2 
    \sqrt{2\big[\epsilon-\Phi(s)\big]} \, ds \ ,
\end{equation}
with $r(\epsilon)$ the inverse function of the gravitational
potential. The overall phase-space volume occupied by energy levels
below $\epsilon$ is thus given by
\begin{eqnarray}
  V & = & \int_{-\infty}^\epsilon\!\! g(\epsilon')d\epsilon' \\
    & = & \frac{2}{3}(4\pi)^2\int_0^{r(\epsilon)} \!\! s^2 
      \left\{2\big[\epsilon-\Phi(s)\big]\right\}^{3/2}\, ds \ .
\end{eqnarray}
Therefore, if $V(f)$ is the cumulative $v(f)$ defined in \Eq{def:VF},
and $f(\Bx,\Bv)$ is a function of the energy alone, we obtain an
integro-differential equation for $f(\epsilon)$:
\begin{equation}
  \label{eq:Vf-integro}
  V[f(\epsilon)] = \frac{2}{3}(4\pi)^2\int_0^{r(\epsilon)} \!\! s^2 
              \left\{2\big[\epsilon-\Phi(s)\big]\right\}^{3/2}\, ds \ .
\end{equation}
This has to be supplemented by the equation that connects $\Phi(r)$ to
$f(\epsilon)$ \citep[][eq. 4-104]{ref:Bin87},
\begin{eqnarray}
  \label{eq:phi-f}
  -\frac{1}{r^2}\frac{d}{dr}\left(r^2 \frac{d\Phi}{dr}\right) 
   & = & 4\pi G\rho(r) \\
  & = & (4\pi)^2 G\int_{\Phi(r)}^0\!\! f(\epsilon)\sqrt{2(\Phi - \epsilon)}\, 
    d\epsilon \ .
\end{eqnarray} 
In general, these equations can be solved numerically, e.g., using a 
Picard iteration scheme. 

In the asymptotic regime, the above equations can be solved
analytically.  If $\rho(r) \propto r^{-\alpha}$ as $r \to 0$, then,
for $\epsilon \to \Phi(0)$, one can show that $f(\epsilon)$ and
$g(\epsilon)$ have the scale-invariant forms
\begin{eqnarray}
  f(\epsilon) &\propto& \big[\epsilon-\phi(0)]
     ^{-\frac{6-\alpha}{2(2-\alpha)}} \ , \\
  g(\epsilon) &\propto& \big[\epsilon-\phi(0)]
     ^{\frac{8-\alpha}{2(2-\alpha)}} \ . 
\label{eq:fg}
\end{eqnarray}
Note that
the derivative of \Eq{eq:Vf-integro} with respect to $\epsilon$ yields 
\begin{equation}
  \label{eq:vf-integro}
  v[f(\epsilon)] = \frac{g(\epsilon)}{f'(\epsilon)} \ .
\end{equation}
By plugging the scale-invariant solutions of \equ{fg} into
\equ{vf-integro}, one finds that for $f\to \infty$, the PDF is also a
power law, $v(f) \propto f^{-\beta}$, with
\begin{equation}
  \label{eq:hi-connection}
  \beta = \frac{18-4\alpha}{6-\alpha} 
  \quad {\rm or} \quad 
    \alpha = \frac{18-6\beta}{4-\beta} \ .
\end{equation}
We learn that the density slopes in the range $0 \leq \alpha \leq 2$,
relevant for the inner regions of haloes where $f$ is high, correspond
to a narrow range of $\beta$ values, $3 \geq \beta \geq 2.5$.  In this
case of power-law profiles, $\beta=2.5$ corresponds to the singular
isothermal sphere $\alpha=2$, $\beta=2.8$ corresponds to an $\alpha=1$
cusp, and $\beta=3$ corresponds to a flat core, $\alpha=0$.

\Eq{eq:hi-connection} can also be obtained from simpler dimensional
arguments using the virial theorem.

\section{MEASURING \boldvf\ IN AN $N$-BODY SYSTEM}
\label{sec:computing}

We wish to measure the phase-space density $f(\Bx,\Bv)$ of a system
represented by $N$ particles of mass $m$ each.  One straightforward
approach would have been to divide phase-space into a large number of
Cartesian cells of equal volume $V$ each.  If cell $j$ contains $N_j$
particles, then the average phase-space density in it can be estimated
as $\bar{f}_j = {N_jm}/{V}$. However, this approach is impractical in
a six-dimensional space. A moderate resolution of 100 bins along each
axis would require $10^{12}$ cells, which is beyond the capacity of
any present-day computer.  Moreover, for the computed $\cf$ to have
any statistical meaning, there should be at least 100 particles in
each cell (for a relative Poisson error of $\sim 10\%$), which
requires a total of $10^{14}$ particles -- about 5 orders of magnitude
beyond the number of particles in today's largest simulations.

A possible way to overcome this problem is by using an \emph{adaptive}
grid, where the cells vary in size to properly allow high resolution
in high-density regions and low resolution in low-density regions.
One can simply create an adaptive grid by dividing each Cartesian cell
which contains more particles than some prescribed threshold $N^*$
into two or more sub-cells. This division can be done recursively
until all cells contain $N^*$ particles or less.

Even more effective would be to vary the shape of the cells as well,
allowing them to adapt more efficiently to the geometry of the
underlying distribution and thus increase the effective resolution.  A
particularly robust method of this type is the \emph{Delaunay
  Tessellation Field Estimator} (DTFE), which has already been
implemented in three dimensions in a cosmological context
\citep[e.g.][ and references therein]{ref:Ber96, ref:Schaap00}.  We
use this method here to measure $f$ in six dimensions.

\subsection{Constructing a Delaunay Tessellation}

A tessellation is the division of $R^d$ space into a complete covering
of mutually disjoint convex polygons.  The Delaunay tessellation
\citep{ref:Del34} is defined for a sample of $N$ points as follows.
The \emph{Delaunay cells} that construct the tessellation are the
$d$-dimensional polyhedrons made by connecting every set of $d+1$
points whose circumsphere [the $(d-1)$-dimensional sphere that passes
through all of them] does not encompass any other point from the
sample.  One immediate advantage of the Delaunay tessellation is that
it is parameter-free, and it completely adapts itself to the
underlying distribution of points. \citet{ref:Schaap00} have
demonstrated the superiority of the DTFE over more conventional
methods for estimating the density in 3D (methods like cloud-in-cell
or the smoothing kernel used in SPH simulations). One may assume that
the same holds for the 6D case.

\begin{figure}
\centerline{ \hbox{
\epsfig{file=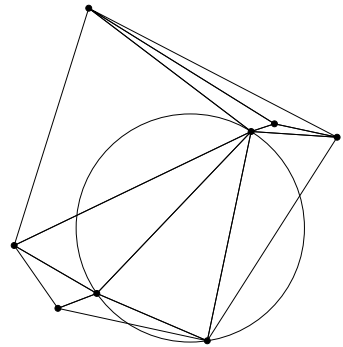,width=0.25\textwidth} } }
\caption{The Delaunay tessellation of 8 points in two dimensions. 
  Shown for example is a circumcircle of one Delaunay cell, which,
by construction, does not encompass any other point.}
\label{fig:simple-tess}
\end{figure}

In 2D, the Delaunay cells are triangles, as illustrated in
\Fig{simple-tess}. In 3D, the Delaunay cells are tetrahedrons. In the
six-dimensional phase space, the Delaunay cells are six-dimensional
polyhedrons, each defined by 7 particles.

\begin{figure}
\centerline{ \hbox{
\epsfig{file=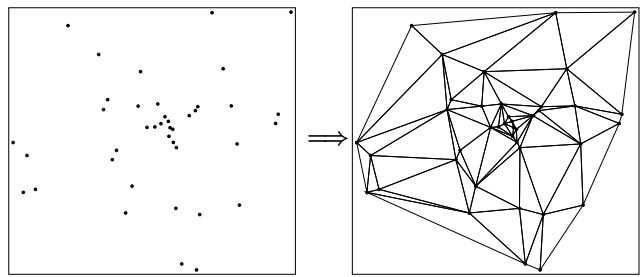,width=0.475\textwidth} } }
\caption{An illustration of the adaptive nature of the Delaunay
  tessellation. Left: an uneven distribution of points in 2D.
Right: the resulting Delaunay tessellation.}
\label{fig:comp-tess}
\end{figure}

\Figu{comp-tess} shows the Delaunay tessellation of an uneven
distribution of points in the 2D plane. It demonstrates the obvious
adaptive nature of the tessellation: regions with high density of
particles are covered by small triangles, whereas regions with low
density are covered with larger triangles.

Constructing the Delaunay tessellation of $N\sim 10^6$ particles in a
six-dimensional space is not a straightforward task. Out of the many
algorithms that exist in the literature \citep[e.g.,][and references
therein]{ref:Su95}, we followed \citet{ref:Ber96} and
\citet{ref:Wey94} in picking the algorithm by \citet{ref:Tan83}.  We
included some of the programming adjustment by van de Weygaert, e.g.,
the use of $k$-trees to speed up searching, and converted the code
from three dimensions to six dimensions.

A full account of the algorithm, the numerical implementation and code
performance, is provided elsewhere (Arad 2004, in preparation).  Here
we just mention briefly that for a halo of $\sim 10^6$ particles the C
code runs for about a week on a common PC with CPU $\sim$2GHz and
internal RAM $\sim$1GB. The resulting tessellation is made of $\sim
10^9$ Delaunay cells, where a typical particle is surrounded by $\sim
7,000$ cells involving $\sim 200$ neighbouring particles.

\subsection{Recovering $f$ and $v(f)$ from the Tessellation}

\begin{figure}
\centerline{ \hbox{ 
\epsfig{file=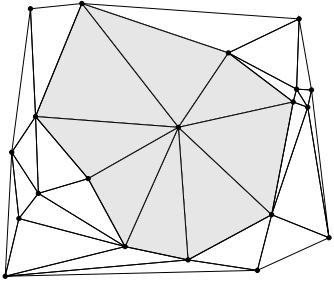,width=0.25\textwidth} }}
\caption{An illustration of a macro cell $W_i$ (grey area),
  the union of the Delaunay cells that surround the particle $i$.}
\label{fig:macro}
\end{figure}

Once the Delaunay tessellation is constructed in phase space, we use
it to estimate $f(\Bx,\Bv)$, generalising the method implemented in 3D
by \citet{ref:Schaap00}.  First we estimate $f_i$ for each particle
$i$.  We define a \emph{macro cell} by joining all its surrounding
Delaunay cells:
\begin{equation}
  \label{def:Wi}
  W_i \equiv \bigcup_\nu D_{\nu_i} \ ,
\end{equation}
where $\{D_j\}$ is the set of all Delaunay cells and $\{D_{\nu_i}\}$
is the subset of cells which contain the particle $i$ as one of their
vertices.  \Figu{macro} shows such a macro cell in 2D. The estimated
phase-space density at point $i$ should be inversely proportional to
the volume of the macro cell, $1/|W_i|$. The proportionality factor
must be greater than unity because the $W_i$ cells of different
particles partly overlap, and one can show that it should in fact be
$d+1$ in order to preserve the total mass of the system.  In
phase-space we therefore define
\begin{equation}
  \label{def:fi}
  f_i \equiv 7\frac{m}{|W_i|} \ .
\end{equation}

In order to estimate $f$ at any general point $(\Bx,\By)$, one might
have used linear interpolation based on the 7 vertices of the
corresponding Delaunay cell $D_\nu$. This means expressing
$f(\Bx,\Bv)$ as a linear function of the 6D vector $\omega = (\Bx,
\Bv)$, namely $f(\omega) = A_\nu \cdot \omega + B_\nu$, with $A_\nu$
and $B_\nu$ a 6D vector and a scalar whose seven unknown values are
found by equating $f(\Bx,\Bv)$ with $f_i$ at the seven vertices of the
cell.  However, with $10^6$ particles involving $\sim 10^9$ Delaunay
cells the linear interpolation is CPU-expensive and practically
impossible. Instead, we perform a zero-order interpolation where
$f(\Bx,\Bv)$ is constant inside each Delaunay cell -- the average of
the $f_i$ values at the seven vertices of $D_\nu$:
\begin{equation}
  \label{eq:f-nu}
  f_\nu \equiv \frac{1}{7}\sum_{i_\nu} f_{i_\nu} \ .
\end{equation}
This has the advantage that any integral of the form
\begin{equation}
  I = \int \!\! d\Bx d\Bv \, \Psi\big[ f(\Bx,\Bv) \big] \ ,
\end{equation}
with $\Psi(\cdot)$ some arbitrary function, can be easily estimated by
the sum
\begin{equation}
  I = \sum_\nu \Psi(f_\nu) |D_\nu| \ .
\end{equation}

In particular, the desired $v(f)$ can be found by first computing
its cumulative counterpart $V(f)$,
\begin{eqnarray}
  V(f_0) &=& \int_{f_0}^\infty \!\! v(f') \, df' 
       = \int_{f(\Bx, \Bv) > f_0}\!\!\!\!\! d\Bx d\Bv \\
   & \to & \sum_{f_\nu > f_0} |D_\nu| \ , 
\end{eqnarray}
and then taking its derivative,
\begin{equation}
  \label{eq:vf-Vf} 
  v(f) = -\frac{dV(f)}{df}  \ .
\end{equation}

In what follows, we often denote the DTFE-measured $f$ and $v(f)$ by
$\fdel$ and $\vdel$ in order to distinguish them from the exact
quantities.

\subsection{Error Estimate}

When we measure the $v(f)$ of a cosmological $N$-body system using the
Delaunay tessellation, we should expect two types of errors.  First
are the errors in the underlying $f$ associated with errors in the
numerical simulation itself, such as errors due to two-body relaxation
effects, force estimation, time integration and so on. A way to
estimate these errors is by re-simulating the same system with
different codes and with different sets of numerical parameters
governing the mass resolution and the force resolution. A systematic
testing of this sort will be reported in an associated paper (Arad,
Dekel \& Stoehr, in preparation).  In the current paper, we make a
preliminary comparison of several different haloes simulated with
different resolutions and with different codes.  We find that all the
virialized haloes tested recover almost the same shape of $v(f)$.

The second type of errors originate from the fact that we try to
estimate a smooth $f(\Bx,\Bv)$ from a finite set of particles using
the Delaunay tessellation. Here we may encounter both statistical and
systematic errors. Some of these errors would decrease as the number
of particles is increased, whereas some are an inherent part of the
DTFE independent of the mass resolution.  In order to obtain a simple
understanding of the nature of the DTFE errors, we first use a simple
statistical model based on a Voronoi tessellation, which resembles the
DTFE but lends itself more easily to analytical treatment.  We then
evaluate the DTFE errors empirically by measuring $\fdel(\Bx,\Bv)$ in
synthetic systems were the particles represent a known $f(\Bx,\Bv)$,
and compare the results to the error-model predictions.  Since this
error analysis is somewhat detached from the main focus of this paper,
we describe it in detail in Appendix~\ref{sec:errors}.  Our wisdom
regarding the DTFE properties and uncertainties can be summarised in
three points as follows:

(i) The measured $\fdel$ at each particle is chosen, to a good extent,
from a probability distribution of the random variable $\fdel/\ftrue$.
In a typical realisation with $10^6$ particles, the width of this
distribution is about one decade, defining the range of fluctuation of
$\fdel$ about $\ftrue$.  As $N\to\infty$, the shape of the
distribution approaches an asymptotic limit $p(\fdel/\ftrue)$ with a
\emph{finite} width of about one quarter of a decade, as deduced from
the DTFE of a Poisson distribution.  Therefore, also in the infinite
limit, the DTFE is expected to produce local fluctuations.

(ii) The measured $\vdel(f)$ can be viewed as a convolution of the
true $v(f)$ and a \emph{fixed window function}, which is related to
the probability distribution $p(\cdot)$ [\Eq{eq:av-vf}]:
\begin{equation}
  \vdel(f=f_0)=\int_0^\infty\!\! \vtrue(f) f^{-1}_0 p(f_0/f)\,df \ .
\end{equation}
For distributions where $v(f)$ is close to a power-law, the difference
between $\vdel(f)$ and $\vtrue(f)$ is negligible over a large range of
scales, as demonstrated in Figs.~\ref{fig:DTFE-vs-exact-c} and
\ref{fig:DTFE-vs-exact-v}.  This is a very useful feature of the
DTFE-measured $v(f)$.

(iii) The relative statistical error in $\vdel (f)$ is proportional to
$1/\sqrt{N}$, and can be approximated by [\Eq{eq:stat-error}]
\begin{equation}
  \label{eq:cal-stat-error}
  \Delta(f) = c\left(\frac{m}{f \left<\Vdel(f)\right>} \right)^{1/2} \ ,
\end{equation}
with $\Vdel(f) = \int_f^\infty \!\!\vdel(f')\,df'$ and $m=M/N$.  The
constant $c$ is of order unity, and can be calibrated by comparing
\Eq{eq:cal-stat-error} to the actual error in lower-resolution
measurements.  In practise, this means that when $N \sim 10^6$ or
more, the statistical error in $\vdel(f)$ is negligible for a very
wide range of $f$. Moreover, in regions where there are large
statistical errors, they are likely to be overwhelmed by systematic
errors.

\section{A UNIVERSAL SCALE-FREE \boldvf}
\label{sec:universal}

We have analysed the $v(f)$ of several different haloes, in three
different mass ranges, simulated within the $\Lambda$CDM cosmology
with two different $N$-body codes.

In order to calculate the $v(f)$ of a given halo, we find the halo
centre using a simple max-density algorithm, and extract all particles
which lie within a distance $R$ from its centre.  The max-density
algorithm is based on an iterative counting in cells: at each
iteration, space is divided into 8 equal cubic cells and the densest
cell is chosen for the following iteration.  The iterative process
stops when the densest cell contains no more than 500 particles, and
the centre of that cell is defined as the centre of the halo.  The
radius $R$ is typically set to be $\sim 10-20\%$ larger than the
virial radius.  We compromise on analysing $\sim 10^6$ particles, such
that we explore a significant dynamical range while the computation
can be completed in a few days.

To estimate the typical phase-space density at the outer regions of
the halo, we define $f_{vir}$ by
\begin{equation}
 f_{vir} = \frac{1}{\pi^{3/2}}\frac{\bar\rho_{vir}}{\sigma^3_{vir}} \ ,
\end{equation}
which is the average space-space density that one would measure for a
halo of constant density $\rho_{vir}$ and a Maxwellian velocity
distribution with a velocity dispersion $\sigma_{vir}$. For a given
halo in a given cosmology, the mean virial quantities $\rho_{vir}$ and
$\sigma_{vir}$ are defined using the virial theorem and the top-hat
model relevant to that cosmology.

As one crude estimate of the upper limit on $f$ for which the
measurement of $v(f)$ is reliable, we evaluate in each halo the $f$
value, $\f02$, below which the \emph{statistical} error in $v(f)$ due
to the DTFE procedure is below $20\%$.  For that we use
\Eq{eq:cal-stat-error}, calibrated by measurements with $10^5$
particles.  This statistical error is expected to be practically
negligible in the range $\fvir < f < \f02$.

\subsection{$N$-body Simulations on Different Scales}

The results described in this paper are based on three different
cosmological simulations. Two of them used the Adaptive Refinement
Tree (ART) code \citep{ref:Kra97}, and the third used the Tree
Particle Mesh (TPM) code \citep{ref:Bode00,ref:Bode03}. In all the
simulations, the assumed cosmological model is the standard
$\Lambda$CDM with $\Omega_{\rm m} = 0.3$, $\Omega_\Lambda = 0.7$ and
$h=0.7$ today.

\begin{figure}
  \centerline{ \hbox{
      \epsfig{file=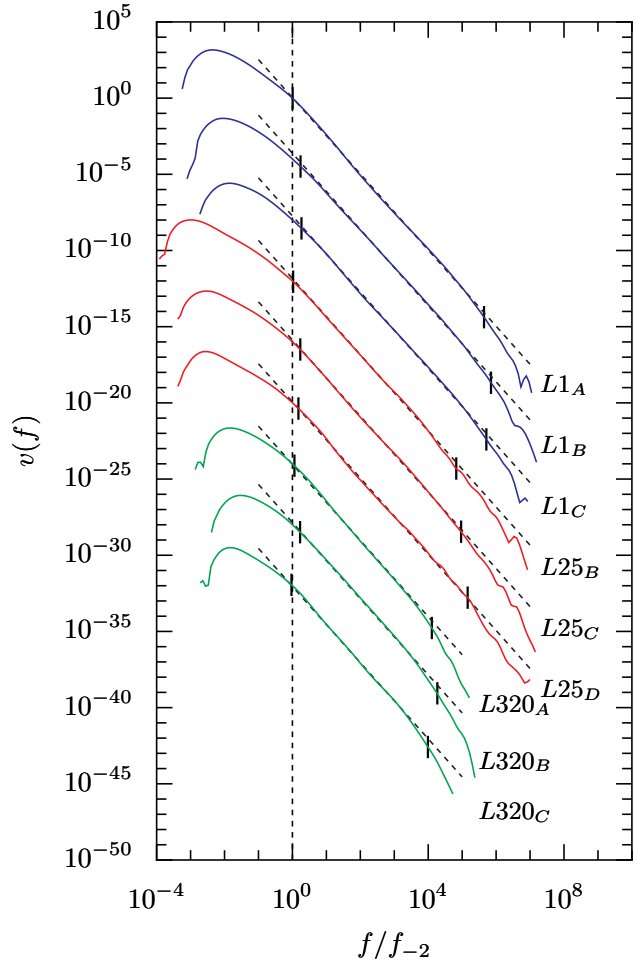,width=0.47\textwidth} }}
\caption{The volume distribution of phase-space density, $v(f)$,
  for each of the nine haloes analysed in this paper
  (see Table~\ref{tab:haloes}). 
  The curves were shifted to coincide at $f=f_{-2}$, where the local
  log slope of $v(f)$ is $-2$, and were then shifted vertically by 4
  decades relative to each other.  A power-law line $v(f)\propto
  f^{-2.5}$ is shown on top of each curve.  Marked on each curve are
  the virial level $\fvir$ and the 20\% statistical error limit
  $\f02$.  The $f$ range is shorter for the L320 haloes because they
  were sampled with less particles.  }
\label{fig:vf}
\end{figure}

The ART simulations were done in periodic boxes of sizes $L=1\hmpc$
and $L=25\hmpc$, whereas the TPM simulation was done in a box of
$L=320\hmpc$. We denote these three simulations by $L1$, $L25$ and
$L320$ respectively. Three haloes were analysed from each of these
simulations, with masses corresponding to to dwarf galaxies
($10^9-10^{10}\msun$), normal galaxies ($\sim 10^{12}\msun$) and
clusters ($\sim 10^{15}\msun$) respectively. Global properties of of
the simulations and the haloes are given in table~\ref{tab:haloes}.
The $v(f)$ curves for these haloes are shown in \Fig{vf}.

\begin{table*}
 \label{tab:haloes}
 \centering
 \begin{minipage}{180mm}
  \caption{Global properties of the 9 haloes analysed in this paper.
    The $v(f)$ of each halo was calculated from all the
    $N_{\mathrm{cut}}$ particles that lie within a radius
    $R_{\mathrm{cut}}$. $\Rvir$ and $\Mvir$ are the the virial radius
    and mass.  $C$ is the concentration of the halo, calculated from
    an NFW fit. $f_2$ is the $f$ level where the logarithmic slope of
    $v(f)$ is $-2$. This value, together with $v(f_2)$, are used to
    scale the $v(f)$ curves of the different haloes in \Fig{vf}.
    Finally, ``Code'', $z$, $m_{\mathrm{par}}$, $r_{\mathrm{force}}$
    and $\sigma_8$ describe the computer code, red-shift,
    mass-resolution, force-resolution, and normalisation of each
    simulation.  
  }
  \begin{tabular}{@{}ccrrcrcccrccr}
  \hline
   Halo     &   $N_{\mathrm{cut}}$  & $R_{\mathrm{cut}}$ & $\Rvir$ & $\Mvir$ &
   $C$ & 
   $f_2$\footnote{units:~$\msun\mbox{Mpc}^{-3}\mbox{km}^{-3}s^{3}$ } & 
   $v(f_2)$\footnote{units:~$\msun^{-1}\mbox{Mpc}^{6}\mbox{km}^{6}s^{-3}$} &
   Code & $z$ & $m_{\mathrm{par}}$ & $r_{\mathrm{force}}$ & $\sigma_8$ \\ 
            &                    &   [kpc]   & [kpc]   & [$\msun$]
   &&&&& & [$\msun$] & [kpc] \\
   \hline\hline
$L1_A$ & $1.6\times 10^6$ & $23$  & $21$  & $1.1\times 10^{10}$ & $7.0$ &
$4.7\times 10^8$ & $1.0 \times 10^{-8}$ & ART & $2.33$ &
$7.0\times 10^3$ & $8.7\times 10^{-2}$ & $0.75$ \\
$L1_B$ & $1.3\times 10^6$ & $23$  & $20$  & $8.6\times 10^{9}$ &  $4.3$ &
$3.4\times 10^8$ & $1.3\times 10^{-8}$ & ART & $2.33$ &
$7.0\times 10^3$ & $8.7\times 10^{-2}$ & $0.75$ \\
$L1_C$ & $1.1\times 10^6$ & $20$  & $19$  & $7.8\times 10^{9}$ &  $7.5$ &
$3.5\times 10^8$ & $1.1\times 10^{-8}$ & ART & $2.33$ &
$7.0\times 10^3$ & $8.7\times 10^{-2}$ & $0.75$ \\
\hline
$L25_B$ & $1.1\times 10^6$ & $420$  & $320$  & $1.9\times 10^{12}$ & $17.4$ &
$5.8\times 10^5$ & $1.3\times 10^0$ & ART & $0$ &
$1.2\times 10^6$ & $1.4\times 10^{-1}$ & $0.9$ \\
$L25_C$ & $1.2\times 10^6$ & $420$  & $330$  & $2.0\times 10^{12}$ & $12.8$ &
$3.4\times 10^5$ & $4.0\times 10^0$ & ART & $0$ &
$1.2\times 10^6$ & $1.4\times 10^{-1}$ & $0.9$ \\
$L25_D$ & $1.4\times 10^6$ & $420$  & $340$  & $2.2\times 10^{12}$ & $11.7$ &
$3.4\times 10^5$ & $4.3\times 10^0$ & ART & $0$ &
$1.2\times 10^6$ & $1.4\times 10^{-1}$ & $0.9$ \\ 
\hline
$L320_A$ & $4.6\times 10^5$ & $3,000$  & $2,700$  & $1.1\times 10^{15}$ & $6.28$ &
$8.8\times 10^2$ & $3.4\times 10^8$ & TPM & $0$ &
$2.6\times 10^9$ & $4.7\times 10^0$ & $0.95$\\ 
$L320_B$ & $4.5\times 10^5$ & $3,000$  & $2,700$  & $1.1\times 10^{15}$ & $5.00$ &
$6.0\times 10^2$ & $6.7\times 10^8$ & TPM & $0$ &
$2.6\times 10^9$ & $4.7\times 10^0$ & $0.95$\\
$L320_C$ & $4.6\times 10^5$ & $3,000$  & $2,700$  & $1.1\times 10^{15}$ & $6.43$ &
$1.0\times 10^3$ & $2.4\times 10^8$ & TPM & $0$ &
$2.6\times 10^9$ & $4.7\times 10^0$ & $0.95$\\
\hline
\end{tabular}
\end{minipage}
\end{table*}

The L1 simulation is from \citet{ref:Col03}.  It is analysed at
$z=2.33$ in \emph{physical} coordinates.  The three haloes studied,
labelled $A$, $B$, $C$, are the largest haloes in the snapshot.  Their
virial radii refer to a mean overdensity of $\Delta=183$, as
appropriate for the given cosmology and redshift. The L25 simulation
is by \citet{ref:Kly01}.  The haloes are denoted $B$, $C$, $D$
following the notation in the simulation paper.  The L320 simulation
is by \citet{ref:Wam04, ref:Wel04}.  The three haloes studied,
labelled $A$, $B$, $C$, are the most massive haloes in the simulation
excluding haloes whose real density map shows an ongoing major merger.

In each of the analysed haloes, we find $v(f)$ to be well described
by a power-law,
\begin{equation}
  v(f) \prop f^{-2.50\pm0.05} \ ,
\end{equation}
over 3 to 5 decades in $f$. It is typically valid between slightly
above $\fvir$ and slightly below $\f02$.  Outside this range, $v(f)$
gradually and systematically deviates downward from the power law.  In
the low-$f$ regime the deviation is associated with departure from the
virial regime, and the high-$f$ deviation is consistent with being due
to the limited mass resolution of the specific halo, as seen by the
proximity to $\f02$ and as demonstrated in Appendix~\ref{sec:errors}.
The high-$f$ deviation from the power-law tends to occur at a smaller
$f$ value in L25, and even smaller in L320, due to the fact that
$\Nvir$ is smaller respectively.

There is no evidence for a significant dependence of $v(f)$ on the
halo mass. There may be a marginal trend for slight steepening of
$v(f)$ as a function of mass, but only from steeper than $f^{-2.45}$
at $\sim 10^{9}\msun$ to flatter than $f^{-2.55}$ at $\sim
10^{15}\msun$.  This indicates relative insensitivity to the exact
slope of the initial fluctuation power spectrum, which varies across
the range from dwarf galaxies to clusters of galaxies. Additionally,
the fact that we obtained essentially the same $v(f)$ from simulations
using two different numerical codes, indicates that the shape of
$v(f)$ is not an artifact of a particular simulation technique.

\section{SUBSTRUCTURES}
\label{sec:substructures}

\subsection{Clumpiness in Phase-Space Density}

Had $f(\Bx,\Bv)$ been a function of the energy alone, and the haloes
were completely spherical and isotropic, the power-law $v(f) \propto
f^{-2.5}$ would have implied via \Eq{eq:hi-connection} that the
real-space density profile must also be a power law, in fact an
isothermal sphere $\rho(r) \propto r^{-2}$, at least over some finite
range in $r$.  This is clearly not the case, as the simulated haloes
are well described by a universal average density profile whose local
logarithmic slope is varying continuously from $-3$ at the outer parts
to $-1$ or even flatter in the inner parts (\se{intro}).  We conclude
that $f$ is far from being a function of energy alone, and in
particular the system must deviate significantly from spherical
symmetry or isotropy.  This could be mostly due to the clumpy
substructure of the halo, where the surviving subhaloes contribute
volume of high phase-space density to $v(f)$, thus making it shallower
than expected from a smooth system with an inner density slope flatter
than $-2$.

\begin{figure}
\centerline{ \hbox{
\epsfig{file=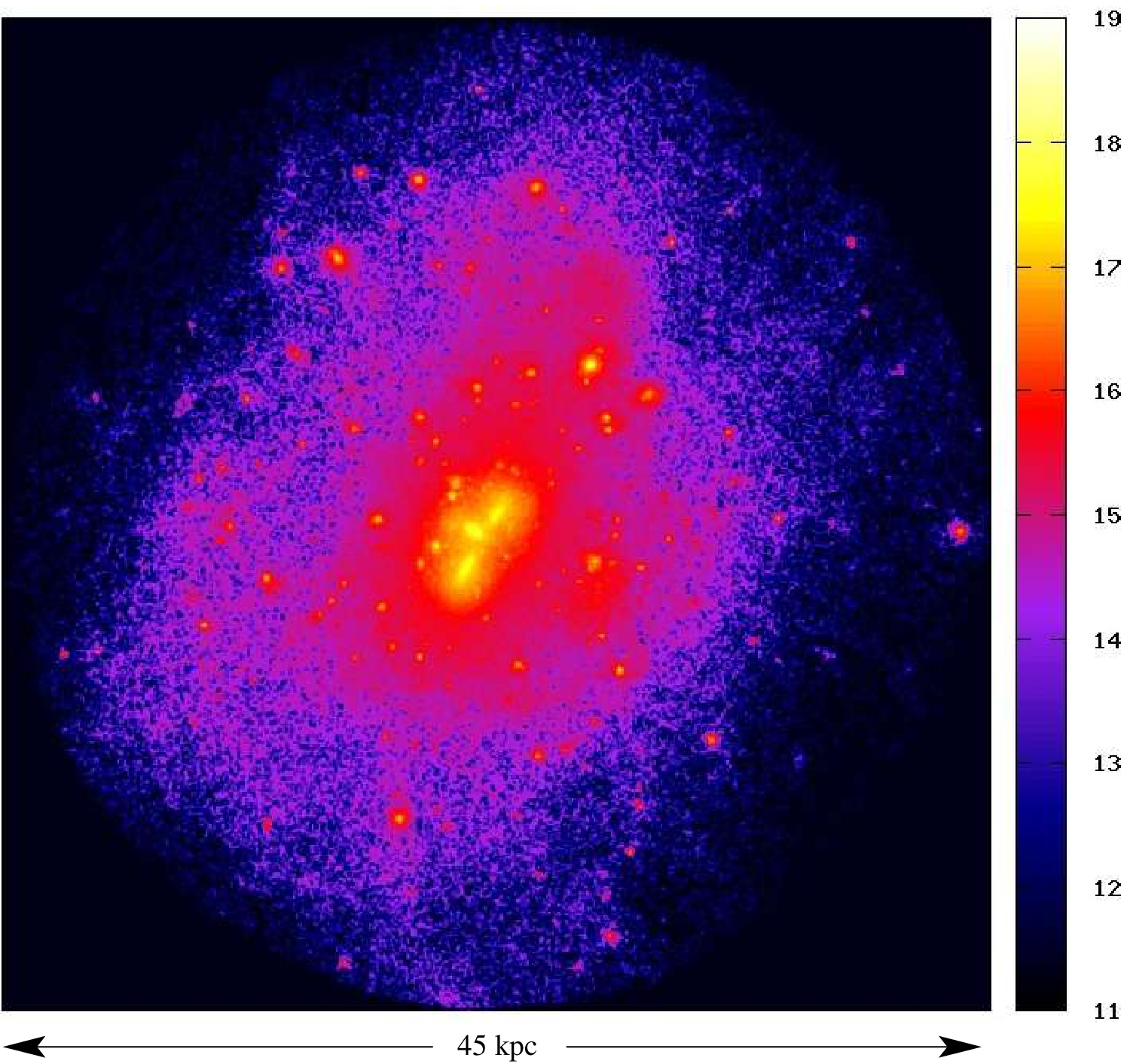,width=0.45\textwidth} } }
\centerline{ \hbox{
\epsfig{file=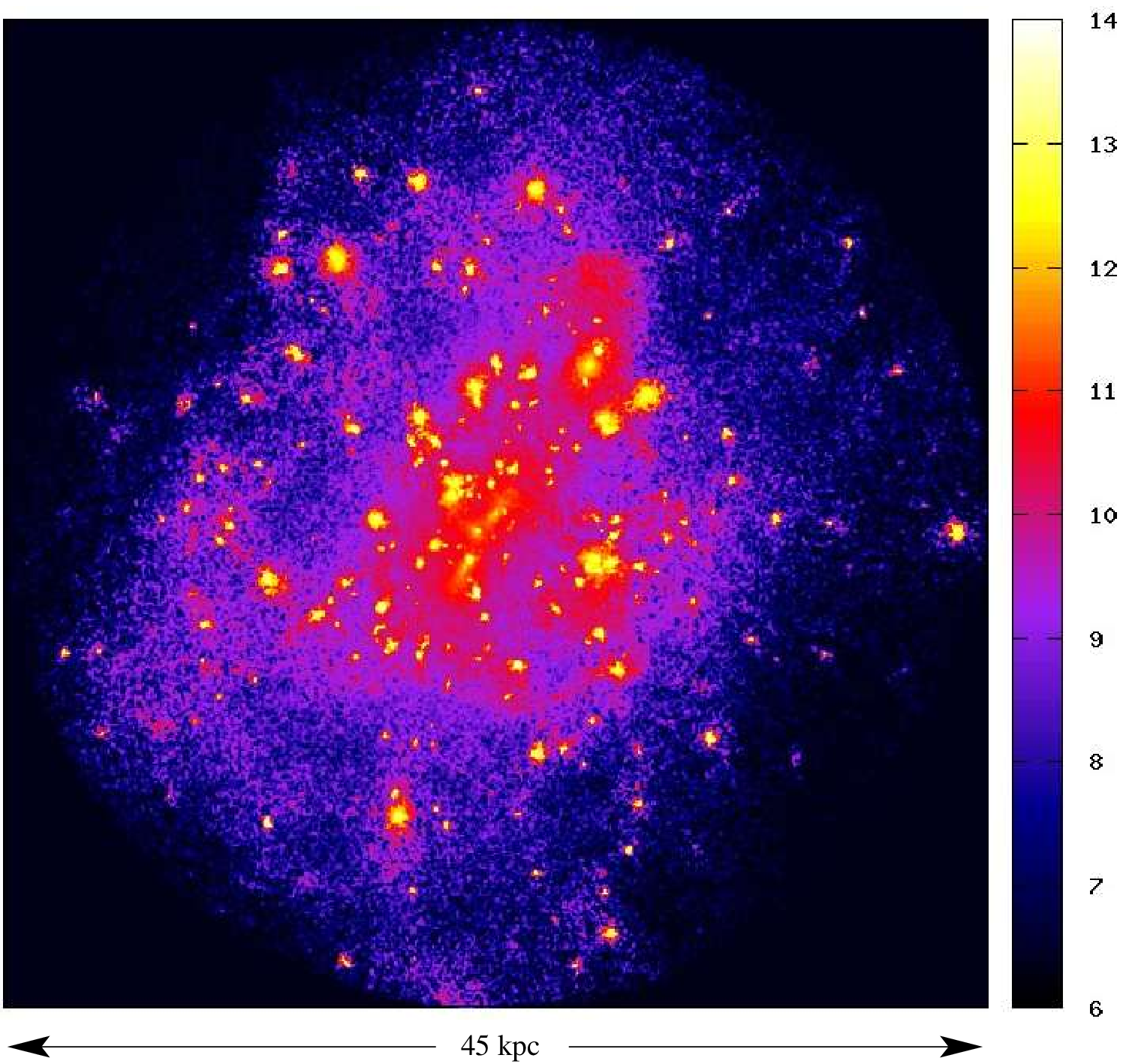,width=0.45\textwidth} } }
\caption{Density maps of dwarf halo $L1_B$ in a slice
  of thickness $0.4 R_{\rm vir}$.  Top: real-space density.  Bottom:
  phase-space density.  The units in the colour key are
  $\log(\rho/[\msun Mpc^{-3}])$ and $\log(f/[\msun
  Mpc^{-3}\mbox{km}^{-3}s^3])$ respectively. The very-high $f$ values
  are found inside clumps which are typically far away from the halo
  centre.}
\label{fig:denmapB}
\end{figure}

\begin{figure}
  \centerline{ \hbox{
      \epsfig{file=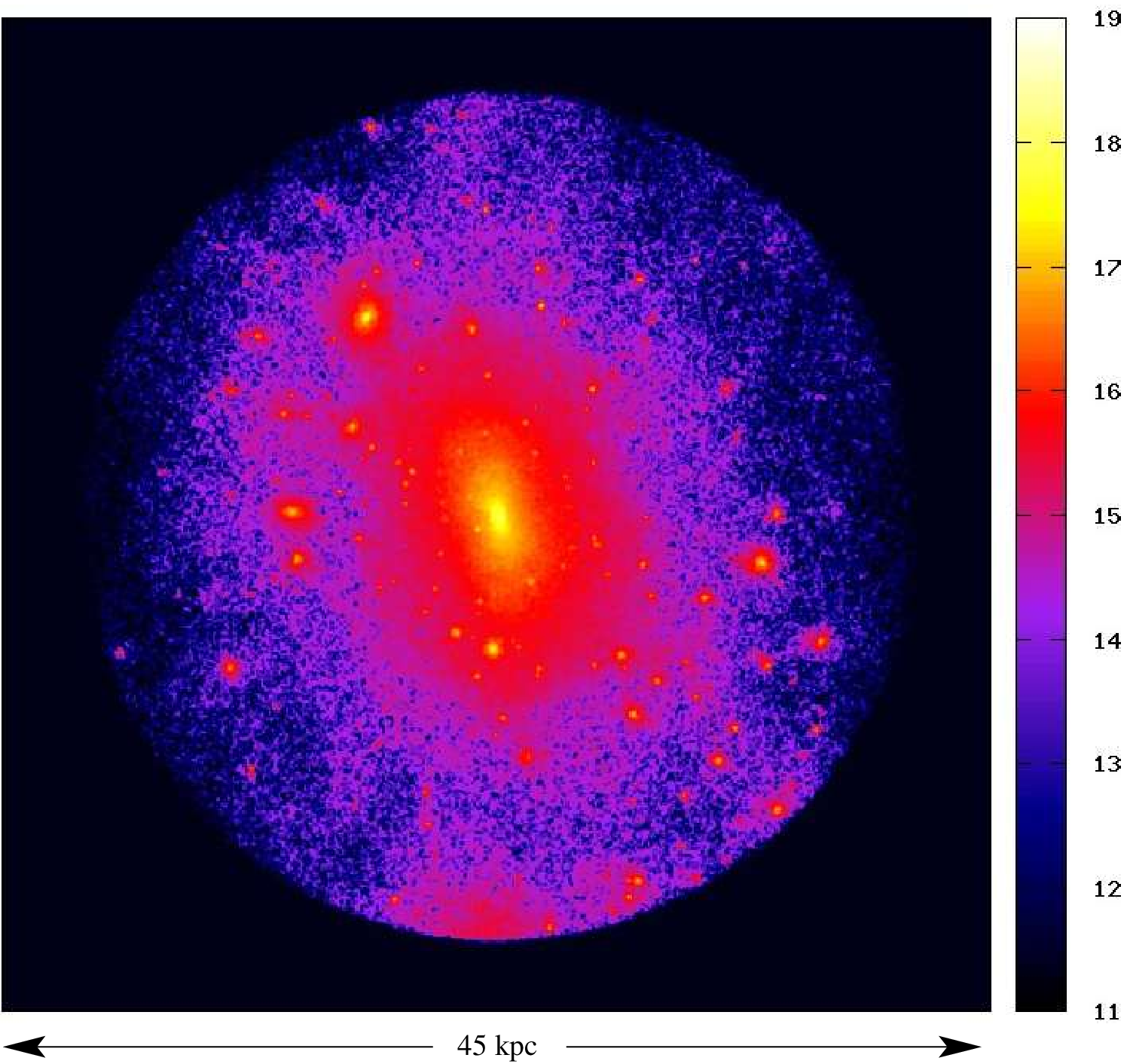,width=0.45\textwidth}
    } } \centerline{ \hbox{
      \epsfig{file=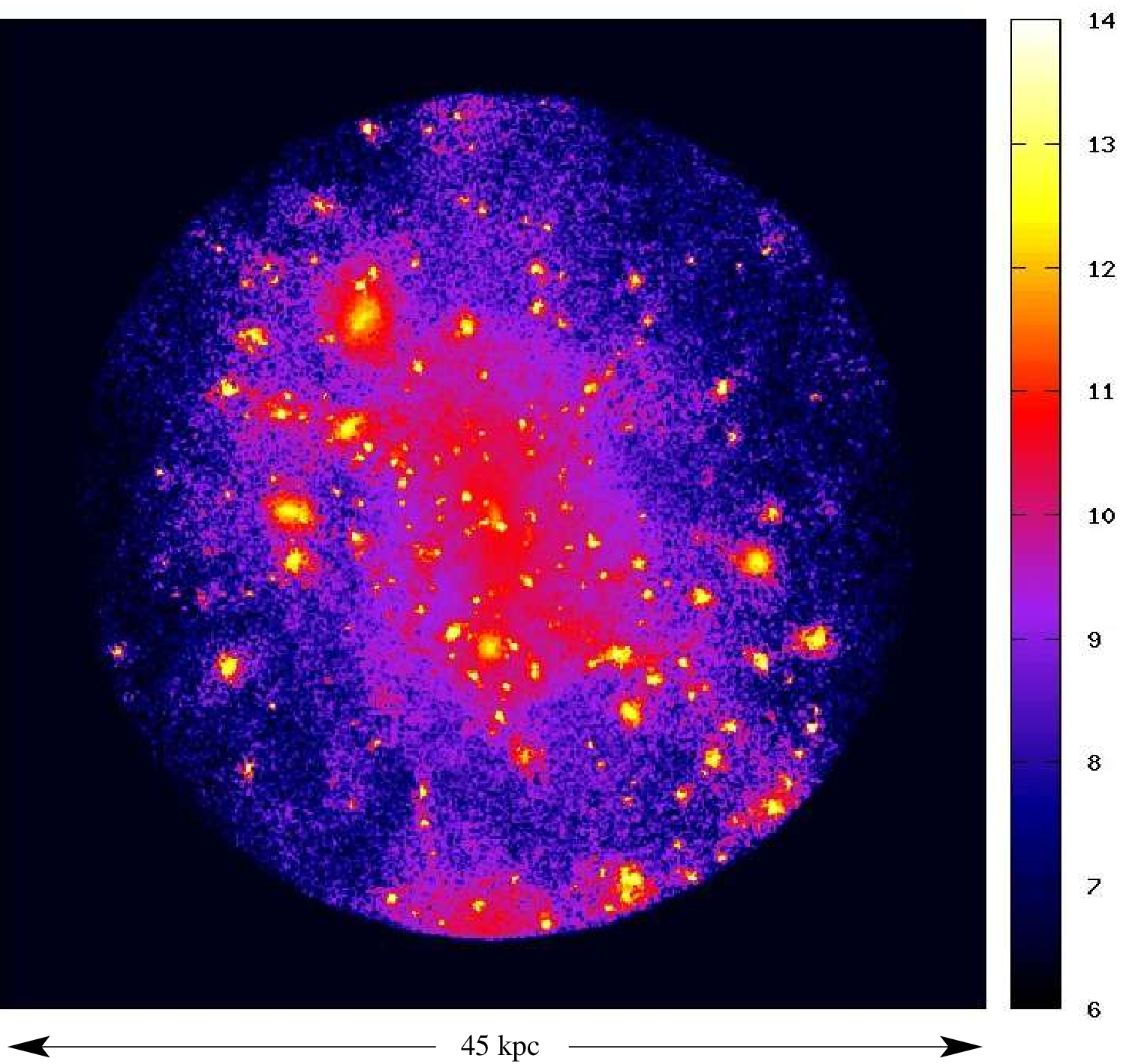,width=0.45\textwidth}
    } }
\caption{Density maps of dwarf halo $L1_C$. See \Fig{denmapB}.}
\label{fig:denmapC}
\end{figure}

In order to address the hypothesis that $v(f)$ is driven by
substructure, we plot real-density and phase-space density maps of
each halo in real-space slices. \Fig{denmapB} and \Fig{denmapC} show
such maps for dwarf haloes B and C (all other haloes show a similar
qualitative behaviour).

The real-space density $\rho$ of each particle was calculated using a
three-dimensional Voronoi tessellation \citep{ref:Wey94}, generated
using the free \texttt{qhull} software package.  We chose this
technique to estimate the real-space density because it is very
similar in its adaptive nature to the Delaunay tessellation technique
used to estimate the six-dimensional phase-space density.  A brief
account of the Voronoi tessellation technique is found in
Appendix~\ref{sec:errors}, where it is used to estimate the errors in
the DTFE method.

The maps were produced in the following way. For each halo, we
consider all the particles within an equatorial slice parallel to the
$xy$-plane, whose width is 40\% of the virial radius.  The slice is
divided into $500 \times 500 \times n_z$ equal cubic cells, with $n_z$
set to have the cells cubic.  The density ($\rho$ or $f$) assigned to
each cell is the average of the densities of all particle within it.
>From each group of cells with the same $x$ and $y$, we plot the one
with the highest density.

We see that the real-space density maps are dominated by the familiar
relatively smooth trend of density decreasing from the centre outward,
with several tight clumps spread throughout the halo. The phase-space
density maps, on the other hand, are qualitatively different.  While
the global trend with radius is much less apparent, the subhaloes
become the highest peaks, especially in the outer regions of the halo.
For example, the clumps with $f> 10^{12}\msun\mpc^{-3}(km/s)^{-3}$
(yellow-reddish colours) are found everywhere.  The very high peaks,
with $f>10^{13}\msun\mpc^{-3}(km/s)^{-3}$ (bright yellow colours), are
all found at a considerable distances from the centre.  The central
peak in $f$ is quite modest in comparison; the elongated structures
near the centre of dwarf-halo B are most likely subhaloes in the
process of merging.

\begin{figure}
  \epsfig{file=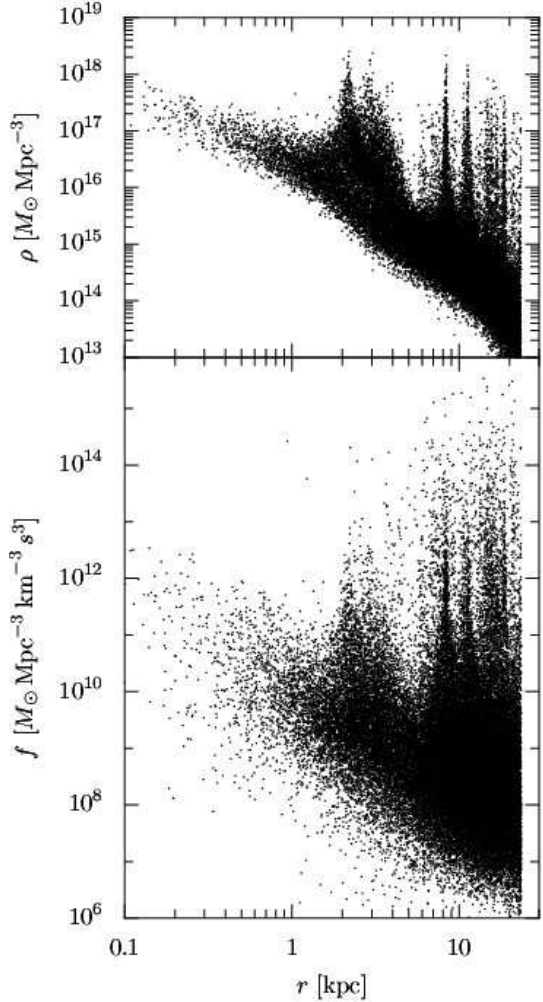,width=0.4\textwidth} 
\caption{Densities as a function of radius for dwarf halo $L1_B$, 
  using a random set of 4\% of the particles.  Top: real-space
  density.  Bottom: phase-space density.  The background particles
  define a general trend of decreasing density with radius, while the
  spikes correspond to subhaloes.  The phase-space density spikes are
  higher than the central peak because the subhaloes are cold. They
  become shorter and broader at smaller radii, indicating heating and
  puffing up by tidal effects and mergers.}
 \label{fig:fr}
\end{figure}

\Fig{fr} highlights the same effect by showing $\rho$ and $f$
associated with a random subset of the N-body particles as a function
of their distance $r$ from the halo centre.  A large portion of these
particles follow the global trend of decreasing density with radius --
they could be associated with a smooth-background component, for which
$f$ is approximately a function of energy alone.  At radii $r>1\kpc$,
the high-$f$ values come in ``spikes" corresponding to the subhaloes.
While the spikes in $\rho$ reach values comparable to the central
peak, the spikes in $f$ could be more than 100 times higher,
indicating that the subhaloes are both \emph{compact} and \emph{cold}.
We note in dwarf halo B, for example, that all the points where
$f\gtrsim 10^{12}\msun\mpc^{-3}\mbox{km}^{-3}\mbox{s}^3$ are in
subhaloes.

The other interesting feature of the spikes is the fact that they seem
to get lower and broader as they get closer to the halo centre, and
that beyond a certain radius (of about $2\kpc$), the spikes completely
blend into the smooth background. This indicates that the subhaloes
phase-mix and lose their high phase-space densities as they approach
the halo centre.  This seems to be the natural result of mergers and
\emph{tidal effects}, which both puff up the subhaloes and heat them
up.

\subsection{Toy Model: Adding Up Small Haloes}

As a first attempt at trying to understand how the power law $v(f)
\propto f^{-2.5}$ in a halo of mass $M$ may originate from its
substructure, we simply add up the typical contributions from the
general population of haloes of different masses $m$ smaller than $M$,
as predicted in the $\Lambda$CDM cosmology.  Based on cosmological
N-body simulations \citep{ref:Mo99a, ref:Ghi00, ref:Luc04}, and in
accordance with the Press-Schechter approximation \citep{ref:Pre74}
and its extensions \citep[e.g., ][]{ref:Lac93, ref:She03}, the mass
function of small-mass haloes (not necessarily subhaloes) can be
approximated by $dn/dm \propto m^{-\gamma}$, with $\gamma \simeq
1.8-2.0$.  Additionally, we assume that the average density profiles
of haloes of different masses have the same functional form and are
simply scaled versions of each other. This is established by $n$-body
simulations for the case of \emph{isolated} haloes, but is less clear
when one considers subhaloes \citep[e.g., ][]{ref:Hay03}.
Nevertheless, we adopt this assumption as our first crude toy model.

As a first approximation we assume that the haloes all form at the
same time in an Einstein-deSitter cosmology, so they have the same
characteristic real-space density $\rho_m=\rho$ (a constant factor
times the universal density), and therefore their typical radii scale
like $r_m \propto m^{1/3}$.  Based on the virial theorem, the velocity
dispersions then scale like $\sigma_m \propto m^{1/3}$. Therefore, the
typical phase-space volume of a halo of mass $m$ scales like $V_m
\propto r_m^3 \sigma_m^3 \propto m^2$, and its typical phase-space
density is $f_m \propto m/V_m \propto m^{-1}$.  If we denote by
$\tilde{v}(f)$ the universal, un-scaled, normalised, dimensionless
probability distribution function relevant for all the haloes, then
the volume distribution function $v_m(f)$ of a halo of mass $m$,
defined such that $v_m(f)df$ is a volume, is given by
\begin{equation}
  v_m(f) = \frac{V_m}{f_m} \tilde{v}\Big(\frac{f}{f_m}\Big) 
         = m^3\tilde{v}(mf) \ .
\end{equation}
Then the total contribution to $v(f)$ of halo $M$ from the
population of smaller haloes $m<\mu M$ is
\begin{equation}
  v(f) = \int_0^{\mu M} \!\! \frac{dn}{dm} v_m(f)\, dm  
             = \int_0^{\mu M} \!\! m^{3-\lambda} \tilde{v}(mf)\, dm
             \ ,
\end{equation}
where $\mu M$ is the mass of the largest subhalo ($\mu <0.5$).
Changing variables $m \to mf$, we finally get
\begin{equation}
  v(f) = f^{-(4-\gamma)}\int_0^{\mu Mf} \!\! 
            s^{3-\gamma} \tilde{v}(s)\, ds \ .
\end{equation}
Thus, $v(f)$ is a multiple of a power-law $f^{-(4-\gamma)}$ and a
monotonically increasing function of $f$; it therefore has to be
shallower than the power law $f^{-(4-\gamma)}$.  For $\gamma \geq
1.8$, this means that $v(f)$ is shallower than $f^{-2.2}$, which is
significantly shallower than the measured $f^{-2.5\pm0.05}$. This
idealised toy model does not seem to work.

This calculation can be generalised to the case where haloes of
smaller masses form first, as implied from the slope of the initial
fluctuation power spectrum $P_k$.  For an Einstein-deSitter cosmology,
the formation time scales with $m$ such that $\rho_m \prop m^{-\nu}$,
with $\nu=(n+3)/2$, and $n$ is the local power index, $n=d\ln P_k/d
\ln k$, at the relevant effective scale for masses $<\mu M$.  Then, in
analogy to the calculation in the previous paragraph, we have
$\sigma_m \prop m^{(2-\nu)/6}$, $V_m \propto m^{(4+\nu)/2}$, $f_m
\prop m^{-(2+\nu)/2}$, and we finally obtain that $v(f)$ must be
flatter than the power law
\begin{equation}
  v(f) \prop f^{-2 -2(2-\gamma)/(2+\nu)} \ .
\end{equation}
For $n=-3$, namely $\nu=0$, the asymptotic index for small haloes
where all haloes form at the same epoch, we recover the earlier result
that $v(f)$ is flatter than $f^{-(4-\gamma)}$.  For $\Lambda$CDM on
galactic scales, the effective power index is $n\leq -2.3$, so $\nu
\leq 0.35$, and for $\gamma \geq 1.8$ we obtain that $v(f)$ is flatter
than $f^{-2.17}$. We see that the time dependence makes only a little
difference.

The total $v(f)$ of the halo should be the sum of the background
contribution and the subhalo contribution, but at the high-$f$ range
we expect $v(f)$ to be dominated by the contribution of subhaloes, as
seen earlier in this section. The above toy model thus predicts that
$v(f)$ should be flatter than $f^{-2.2}$, in conflict with the
measured $v(f) \prop f^{-2.5}$.

We conclude that a halo is not simply an ensemble of clumps drawn from
the general population of smaller haloes.  The subhaloes may have a
different mass function, their shape properties may vary differently
with mass, and they both could vary with distance $r$ from the
host-halo centre. If one keeps the scaling relation with $\nu=0$ and
ignores any variation with $r$, the required subhalo mass function for
matching the measured $v(f) \prop f^{-2.5}$ is with $\gamma = 1.5$
(compared to $\gamma=1.8$). Indeed, a flattening of the subhalo mass
function could be a natural result of the inevitable dynamical
evolution of the subhaloes in the potential well of their host halo.
The phase mixing due to tidal effects, including total disruption, is
likely to be more efficient in less massive subhaloes, thus flattening
the mass function. Also, dynamical friction is more efficient in
making more massive subhaloes sink into the halo, thus making the mass
function flatter in the inner parts.  However, recent simulations
indicate that the subhalo mass function is not flatter than $\gamma
\simeq 1.7$ \citep{ref:Luc04}, indicating that the tidal effects on
the inner structure of subhaloes must also have an important role.
These are matters for more detailed future studies, but the failure of
the idealised toy model analysed above to reproduce the magic power
law $v(f) \prop f^{-2.5}$ indicates that the phase-space density is
likely to provide a useful tool for studying the dynamical evolution
of subhaloes in host haloes.

\section{DISCUSSION AND CONCLUSION}
\label{sec:conc}

Using Delaunay tessellation, we developed a method for measuring the
6-dimensional coarse-grained phase-space density $f(\Bx,\Bv)$ in
$N$-body systems.  We focused, in particular, on measuring the
phase-space volume distribution function, $v(f)$.  We applied this
technique to several simulated haloes of $\sim 10^6$ particles, formed
by hierarchical clustering in the standard $\Lambda$CDM scenario, and
obtained two striking new results.

First, $v(f)$ is well described by a power law, $v(f) \propto f^{-2.5
  \pm 0.05}$, over 3 to 5 decades in $f$.  The power-law regime starts
at an $f$ value which corresponds to the characteristic size of the
virialized halo.  It ends at an $f$ value which is determined by the
dynamical resolution limit of the specific simulation.  Therefore, the
true power-law range may extend to $f \rightarrow \infty$.  This power
law seems to be insensitive to the halo mass in the range
$10^9-10^{15}\msun$, indicating insensitivity to the exact slope of
the fluctuation power spectrum, as long as the haloes are built by
hierarchical merging of clumps bottom up.

Second, this power-law originates from substructures within the halo
rather then the overall trend with radius. The substructure completely
dominates the high-$f$ parts of the $v(f)$ distribution.  The
infalling clumps seem to phase-mix --- by puffing up, heating and
stripping --- as their orbits decay from the virial radius inwards
toward the halo centre and they melt into the halo smooth background.

Our first worry is that these results could be numerical artifacts, or
severely contaminated by such.  Based on our error analysis and tests
with mock datasets, we argue that the $v(f)$ measured by the DTFE
algorithm genuinely reflects the true phase-space properties of the
given $N$-body system over a broad range of $f$.  The question is
whether the phase mixing suffered by the subclumps as they approach
the halo centre might be an artifact of numerical effects such as
two-body relaxation, leading to underestimated inner densities and/or
overestimated internal velocities.  A similar effect has been pointed
out using a one-dimensional toy model \citep{ref:Bin03}.  The apparent
agreement between simulations run with different codes and different
resolutions is encouraging.  In order to specifically address the
effect of few-body relaxation, we intend to run twice a simulation of
the same halo with the same number of particles but with a different
force resolution (ongoing work with F. Stoehr).

Assuming that the simulations genuinely reflect the true physical
behaviour under the Vlasov equation, the origin of the robust
power-law shape of $v(f)$ from the merging substructure becomes a very
interesting theoretical issue.  As demonstrated in \se{substructures},
a simple model using the mass function and the scaled profiles of the
general halo population in the $\Lambda$CDM scenario does not
reproduce the correct power law.  This, and the apparent trend of the
$f$ spikes with radius, indicate that the structural and kinematic
evolution of the subhaloes in the parent halo are important. Studies
of tidal heating and stripping may be found useful in this modelling.

It would be interesting to follow the phase-space evolution and the
contribution to the overall $v(f)$ by a single, highly resolved
subhalo, or many of those, as they orbit within the parent halo and
approach its centre. This may help us understand the nature of the
interaction between the parent halo and its subhaloes, and the origin
of the $v(f)$ power law (ongoing works with E. Hayashi and with B.
Moore).

Another more general but speculative possibility is that the
$f^{-2.5}$ power law represents some sort of a cascade of relaxation
processed in phase-space, in which high phase-space densities turn
into lower (coarse-grained) densities through the process of mixing.
In general, the fact that our findings are expressed in terms of the
fundamental concept of phase-space density should make them more
directly accessible to analytical treatment. In this respect, it may
prove beneficial to investigate more closely the time evolution of the
$v(f)$ of a cosmological halo and its components.  This may shed light
on the connection between the $v(f)$ power-law behaviour and the
relaxation processes within the halo.

We saw that the power-law behavior of $v(f)$ is limited to the virial
regime.  It would be interesting to learn how this shape evolves in
time as the halo virializes.  A preliminary study (to be concluded and
reported in another paper) indicates that in the intermediate-$f$
regime the $v(f)$ of a pre-virialized system is significantly flatter
than $f^{-2.5}$, while in the high-$f$ regime it drops in a much
steeper way.  The $f^{-2.5}$ behaviour seems to be a feature unique to
virialized systems.

We learnt that in the haloes that are built by hierarchical
clustering, the power-law behaviour $v(f)\propto f^{-2.5}$ reflects the
halo substructure. It would be interesting to find out whether this
power-law behaviour actually requires substructure, or it is a more
general phenomenon of virialized gravitating systems, valid
independently of substructure.  One way to answer this question would
be to analyse simulated haloes in which all fluctuations of
wavelengths smaller than the halo scale were removed, resulting in a
smooth halo formed by monolithic collapse, with no apparent
substructure in the final configuration.  As described in \se{intro},
such haloes are known to still have NFW-like density profiles in real
space, and one wonders whether they also have the magic power-law
$v(f)$.  There are preliminary indications for a steeper $v(f)$ in
this case (Arad, Dekel \& Moore, in preparation).  If confirmed, it
would indicate that the $f^{-2.5}$ behaviour, while insensitive to the
exact slope of the initial power spectrum, is unique to the
hierarchical clustering process, and is not a general result of
violent relaxation.

Our current results are just first hints from what seems to be a
promising rich new tool for analysing the dynamics and structure of
virialized gravitating systems. The analysis could become even more
interesting when applied to haloes including the associated gaseous
and stellar components.

\section*{Acknowledgments}
We thank Paul Bode for making his earlier simulations available for
analysis.  We acknowledge related ongoing collaborations with Stefan
Gottloeber, Eric Hayashi \& Julio Navarro, Ben Moore and Felix Stoehr.
We are grateful for stimulating discussions with Stefan Colombi,
Donald Lynden-Bell, Gary Mamon, Jerry Ostriker and Simon White.  Part
of this work has been done at the IAP in Paris.  AK acknowledges
support of NASA and ASF grants to NMSU. Our simulations were performed
at the National Energy Research Scientific Computing Centre (NERSC).
IA is a Marie Curie Postdoctoral Fellow.  AD is a Miller Visiting
Professor at UC Berkeley. This research has been supported by the
Israel Science Foundation grant 213/02, by the German-Israel Science
Foundation grant I-629-62.14/1999, and by NASA ATP grant NAG5-8218.


\appendix
\section{ERRORS IN \boldvdel}
\label{sec:errors}

The ``true", underlying coarse-grained $f(\Bx,\Bv)$ is what one would
have measured using a fixed smoothing window in phase space (e.g.
counts in fixed cells) and an arbitrarily large number of particles.
Instead, the Delaunay Tessellation Field Estimator uses adaptive cells
in order to deal with the mass resolution limitations.  Therefore, the
relation between the measured $\vdel(f)$ and the underlying
$\vtrue(f)$ is not trivial.  Both statistical and systematic errors
might influence our results.  We estimate these errors empirically
below, but we start with an approximate model that provides a simple
understanding of the origin of the errors.

\subsection{A Voronoi Model: Fixed Smoothing $v(f)$}

A technique similar to the Delaunay tessellation, but somewhat simpler
to interpret, is the Voronoi tessellation \citep{ref:Wey94}.  For each
particle, the Voronoi cell is defined as the region of phase space in
which every point is closer to that particle than to any other
particle.  In this case, $N$ particles define exactly $N$ Voronoi
cells which cover all of phase space with no overlaps.  If $V_i$ is
the Voronoi cell of particle $i$, then a natural mass-preserving way
of estimating the phase-space density inside that cell is by $f_i
\equiv m_i/|V_i|$. We denote the quantities measured this way by
$\fvor$ and $\vvor(f)$.

Much like the Delaunay tessellation, the Voronoi tessellation is an
adaptive grid that enables one to estimate $f(\Bx,\Bv)$ even in the
presence of a relatively small number of particles.  We use the
Delaunay method in our main analysis because it is somewhat more
accurate \citep{ref:Schaap00}, and is easier to calculate numerically.
However, the similar Voronoi method provides a simple way of learning
about the properties of the measured $v(f)$ and understanding the
uncertainties associated with such a measurement. The empirical tests
of the Delaunay measurements below demonstrate the relevance of the
wisdom gained by analysing the Voronoi model.

To understand the errors in the Voronoi density estimation, let us
start with the trivial case where all of infinite phase-space is
uniformly filled with phase-space density $f_0$, which is represented
by an infinite number of particles with mass $m$.  A volume $V$ of
phase-space would then contain on average a finite number of
particles, $f_0 V/m$. The Voronoi estimate of $f_i$ for each particle
would fluctuate about $f_0$ due to the discreteness of the particle
distribution.  Since there is no typical scale in the problem, and
each cell always contains one particle, the fluctuations $\delta f/f$
per Voronoi cell would remain at the same level even if one increases
the average number density of particles while decreasing the mass of
each particle in proportion, keeping $f_0$ the same.  Therefore, the
probability that the Voronoi estimated $f$ would lie in the interval
$f\to f+df$ may be written in terms of a universal probability
distribution function $p_\infty(f/f_0)\,d(f/f_0)$.

When we consider a finite system in a box of volume $V$, with a finite
number of particles $N$ inside it, we may expect the probability
distribution $p_N(\cdot)$ to deviate from its asymptotic form
$p_\infty(\cdot)$.  As $N$ decreases, we expect $p_N(\cdot)$ to widen
due to the increasing effect of the boundaries.  Next, examine a
system with a non-uniform phase-space density, such as a cosmological
halo. If the number of particles that represent this $f$ is
sufficiently large, we may approximate every region in phase-space as
being locally uniform, and estimate its Voronoi $f$ using the
asymptotic $p_\infty(\cdot)$. This assumption is expected to break
down in regions with very high phase-space density, where the sampling
may become poor and insufficient, or in regions where $f$ has large
gradients over small scales.  Nevertheless, lets assume for the moment
that there exists an effective $p(\cdot)$ [not necessarily
$p_\infty(\cdot)$] which properly approximates the fluctuation
distribution of the Voronoi $f$ for all particles.
 
This assumption allows us to calculate the expectation value of
$\vvor(f)$ for a finite system with a given $f(\Bx,\Bv)$.  This is
done by first calculating the average of $\Vvor(f)$, the cumulative
version of $\vvor(f)$ defined in~(\ref{def:VF}), and then
differentiate it to obtain the average of $\vvor(f)$. Assuming that
$f(\Bx,\Bv)$ is realised by $N\gg 1$ particles, we divide phase space
into a large number of cells $\omega_\alpha$, which are small enough
to guarantee that (a) each cell is very unlikely to contain more than
one of the $N$ particles, and (b) the value of $f(\Bx,\Bv)$ is
approximately constant within every cell.

For each cell $\omega_\alpha$, we calculate
$\left<V_{\alpha}(f)\right>$, the average of the contribution of this
cell to $\Vvor(f)$. The contribution $V_{\alpha}(f)$ would be non-zero
only when there is a particle in the cell $\omega_\alpha$, a particle
whose assigned Voronoi estimated density is $\fvor$.  If $f_\alpha$ is
the true phase-space density in that the cell, then, according to our
assumption, $\fvor$ would be chosen at random out of the probability
distribution $p(\fvor/f_\alpha)\, d(\fvor/f_{\alpha})$.  Once $\fvor$
is chosen, $V_\alpha(f)$ is given by
\begin{equation}
  V_\alpha(f) = \left\{ 
      \begin{array}{lcl}
        m/\fvor &,& f<\fvor \\
        0         &,& f>\fvor
      \end{array} 
    \right. \ . 
\end{equation}
Using our assumption (a) above, the probability of $\omega_\alpha$ to
host one particle is $P_\alpha = NM^{-1}\int_{\omega_\alpha}
\!\!f(\Bx,\Bv)\, d\Bx d\Bv$, with $M=Nm$
the total mass of the system. Therefore,
\begin{eqnarray}
  \left< V_\alpha(f)\right> &=& P_\alpha \times \int_{f}^\infty 
    p(\fvor/f_\alpha) d(\fvor/f_{\alpha}) \, m/\fvor \\
  &=& \int_{\omega_\alpha} d\Bx d\Bv \int_{f}^\infty 
    p[\fvor/f(\Bx,\Bv)]  f^{-1}_{vor}\, d\fvor \ ,
\end{eqnarray}
and so,
\begin{eqnarray}
 \label{def:V-alpha}
  \left< \Vvor(f)\right> &=& \sum_\alpha \left<
       V_\alpha(f)\right> \nonumber \\
  &=& \int\!\! d\Bx d\Bv \int_{f}^\infty 
    p[\fvor/f(\Bx,\Bv)]  f^{-1}_{vor}\, d\fvor \nonumber \\
  &=& \int_0^\infty\!\!df'\, v(f') \int_{f}^\infty 
    p(\fvor/f')  f^{-1}_{vor}\, d\fvor \ .
  \label{eq:av-VF}
\end{eqnarray}
In the last equality, we have used the exact $v(f)$ to replace the
six-dimensional phase-space integration. By differentiating
\Eq{eq:av-VF} with respect to $f_0$, we finally obtain the desired
expectation value:
\begin{equation}
  \label{eq:av-vf}
  \left< \vvor(f)\right> = 
    \int_0^\infty\!\! v(f') f'^{-1} p(f/f')\,df'   \ .
\end{equation}

We see that the measured $\left<\vvor(f)\right>$ is a convolution of
the exact $v(f)$ with a \emph{fixed window function}
$p(f/f_{\ftrue})$.  The narrower $p(\cdot)$ is, the closer
$\left<\vvor(f)\right>$ would be to the true $v(f)$. However, as
argued above, even when $N\to\infty$ the window $p(\cdot)$ does not
approach a Dirac delta function; it rather converges to some
finite-width distribution $p_\infty(\cdot)$.  Therefore, even with an
infinite resolution the Voronoi tessellation would not produce the
exact $v(f)$; it converges to a convolution of it with a fixed window
$p_\infty(\cdot)$.

This convolution would not affect the shape of the measured
$\left<\vvor(f)\right>$, and would preserve the true shape of $v(f)$,
provided that $v(f)$ does not vary drastically over $f$ scales which
are smaller than the width of the window.  In particular, when $v(f)$
is a power law, the Voronoi algorithm would recover the same power law
for $\left<\vvor(f_0)\right>$.

\subsection{Empirical Testing with Mock Systems}

For an empirical study of the errors in the DTFE-measured $v(f)$, and
for testing how well the Voronoi model approximates these errors, we
have performed a series of $\vdel(f)$ measurements on systems with
known phase-space densities of the form
\begin{equation}
  \label{eq:f-maxwell}
  f(\Bx,\Bv) = \rho(x) \big[2\pi
  \sigma^2(x)\big]^{-3/2}e^{-v^2/2\sigma^2(x)} \ ,
\end{equation}
corresponding to a spherical system in real space with a Maxwellian velocity
dispersion. We have examined six such systems with three different
density profiles parametrised by $\alpha$,
\begin{equation}
  \label{eq:syn-rho}
  \rho_\alpha(x) = \frac{e^{-x/5}}{x^\alpha(1+x)^{3-\alpha}} \ , 
  \quad \alpha = 0,\ 0.5,\ 1.0 \ ,
\end{equation}
and the following two types of dispersion profiles:
\begin{eqnarray}
  \sigv(x) &=& \left[ \frac{M(x)}{x}\right]^{1/2} \ , \\
  \sigc(x) &=& 0.1 \ .
\end{eqnarray}
The subscripts ``v'' or ``c'' denote a varying dispersion profile versus
a constant one respectively. 
The $v(f)$ for such systems is
\begin{equation}
  v(f) = \frac{(4\pi)^2}{f} \int_0^{x(f)}\!\! x^2 \sigma^3(x) 
     \sqrt{2\log\frac{f(x)}{f}}\, dx \ ,
\end{equation}
with 
\begin{equation}
  f(x) \equiv \frac{\rho(x)}{\big[2\pi\sigma^2(x)\big]^{3/2}} \ ,
\end{equation}
and $x(f)$ its inverse.

\begin{figure}
  \epsfig{file=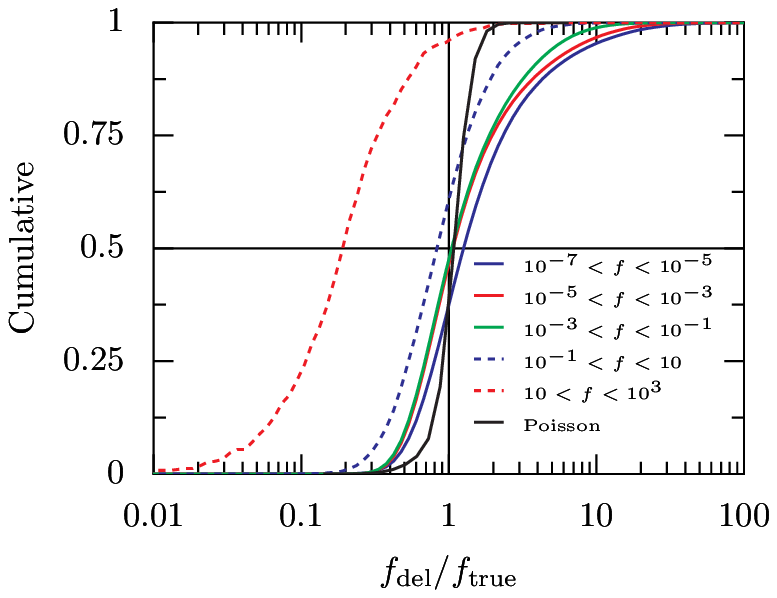,width=0.45\textwidth}
  \epsfig{file=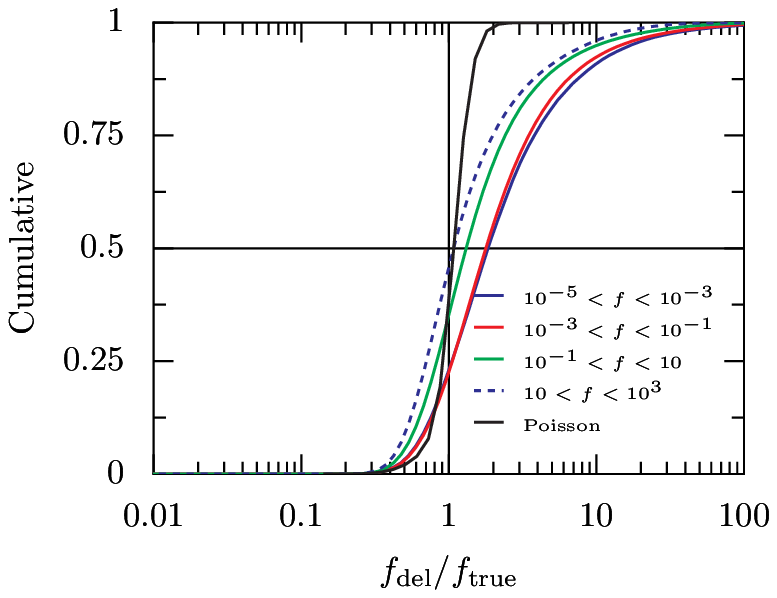,width=0.45\textwidth}
\caption{
  The cumulative distribution of $\fdel/\ftrue$ as measured in
  different bins of $\ftrue$ for two mock systems described by
  \Eq{eq:f-maxwell} with $10^6$ particles.  Top: an NFW $\rho(x)$
  given by \Eq{eq:syn-rho} ($\alpha=1.0$) with a varying $\sigv^2(x) =
  M(x)/x$.  Bottom: same $\rho(x)$ but with a constant $\sigc(x) =
  0.1$.  The solid black line represents the cumulative distribution
  of $\fdel/\ftrue$ for a homogeneous Poisson distribution, which is
  realised in a six-dimensional cube using $10^6$ particles. This
  should be regarded as the asymptotic limit with an infinite
  resolution.
}
\label{fig:f-bins}
\end{figure}

Figure~\ref{fig:f-bins} shows the cumulative distributions of
$\fdel/\ftrue$ in different bins $(f_j,f_{j+1})$ of $\ftrue$, for the
$\alpha=1$ $\sigv$ system and the $\alpha=1$ $\sigc$ system.  Both
systems were realised using $10^6$ particles. In both cases we have
also plotted the cumulative distribution $\fdel/\ftrue$ for a
homogeneous Poisson distribution, realised in a six-dimensional cubic
box with $10^6$ particles. We have verified that this distribution is
essentially unchanged when the calculation is done with $10^5$
particles. Therefore, it should be regarded as good approximation to
the asymptotic limit we would have reached in the different bins, had
we used an infinite number of particles. The other four systems give
essentially the same results.

In the 1-v system, five bins were defined by $f_j=10^{-7}, 10^{-5},
10^{-3}, 10^{-1}, 10, 10^{3}$. We see that the shape of the
distribution, corresponding to the width of the differential
distribution, is very much independent of $\ftrue$. For all bins, the
full width is less than a 1.5 decades. On the other hand, there does
seem to be a systematic shift of the median toward larger values of
$\fdel/\ftrue$ for smaller values of $\ftrue$. The large shift of the
highest-$f$ bin ($10<f<10^3$) toward small $\fdel/\ftrue$ is a result
of the very low number of particles in that bin, less than 500, which
is insufficient for representing such high phase-space densities.

The bias toward larger phase-space densities in the low-$f_{\ftrue}$
bins may be attributed to boundary effects: while the total mass of
the $N$ particles in the realisation is equal to the total mass one
would obtain from the exact $f(\Bx,\Bv)$ integrated over the infinite
phase-space, the total phase-space volume used by the DTFE to estimate
$f(\Bx,\Bv)$ is finite. It is the smallest possible convex polygon
containing all $N$ particles. Therefore, we may expect an overestimate
of $f$, which would be more pronounced near the boundaries.
Nevertheless, as we shall see by comparing $\vdel(f)$ to $\vtrue(f)$,
on scales of a few-decades, this bias is rather meaningless.

The $\fdel/\ftrue$ distributions of the 1-c system are essentially the
same as the ones in the 1-v system. Here 4 bins were defined by
$f_i=10^{-5}, 10^{-3}, 10^{-1}, 10^1, 10^3$. The overall shape of the
plots changes very little from bin to bin, and its width is about a
decade and a half. Additionally there exists the bias towards larger
$\fdel/\ftrue$ ratios as $\ftrue$ gets smaller.

\begin{figure}
\centerline{ \hbox{
\epsfig{file=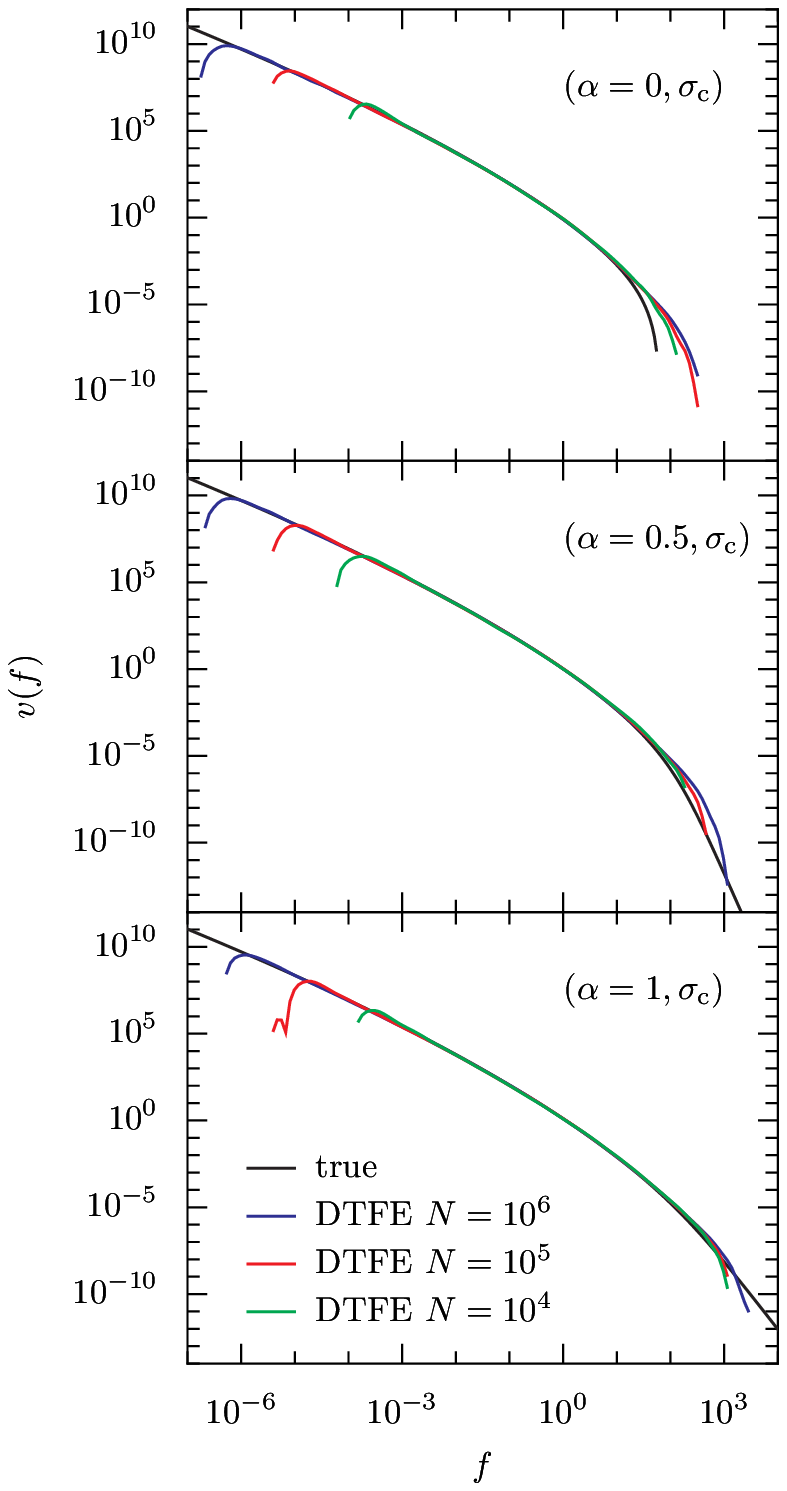,width=0.45\textwidth}}}
\caption{The measured $\vdel(f)$ versus the true $v(f)$ for the three
  $\sigc$ mock systems, each performed with three different resolutions.}
\label{fig:DTFE-vs-exact-c}
\end{figure}

\begin{figure}
\centerline{ \hbox{
\epsfig{file=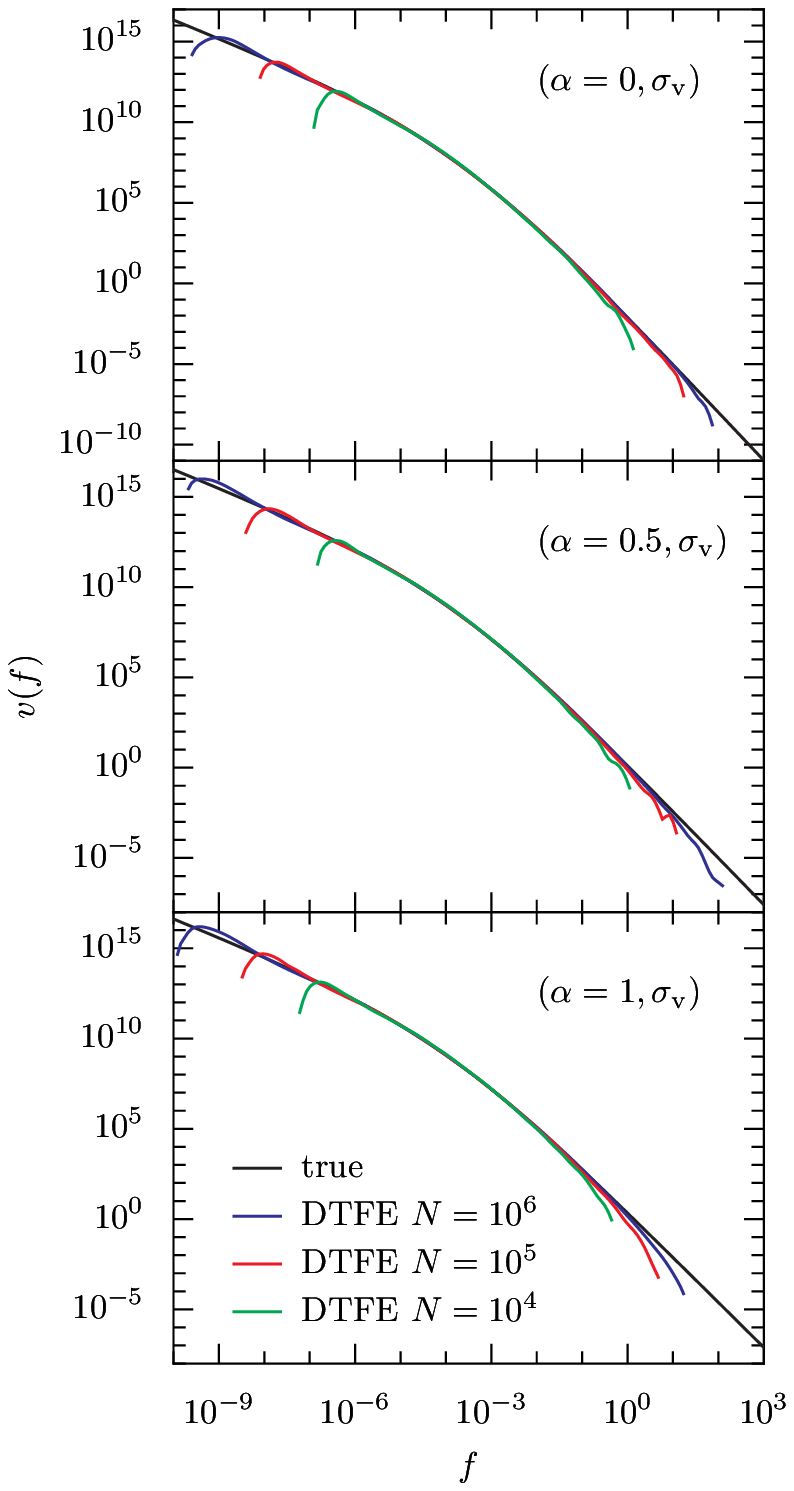,width=0.45\textwidth}}}
\caption{Same as \Fig{DTFE-vs-exact-c} but for the
  $\sigv$ mock systems.}
\label{fig:DTFE-vs-exact-v}
\end{figure}

To see how well the DTFE reconstructs the true $v(f)$, we have
measured the $\vdel(f)$ of the six mock systems using $10^4$, $10^5$
and $10^6$ particles for each system.
Figures~\ref{fig:DTFE-vs-exact-c},~\ref{fig:DTFE-vs-exact-v} present
the results of these numerical experiments. We see that with the
highest resolution, of $10^6$ particles, the recovery is excellent
over a range of almost 7 decades in $f$. Systematic deviations begin
at the high-$f$ end and the low-$f$ end. At both ends, the deviations
appear at about one to two decades inward to the highest and lowest
values of $\fdel$ in that realisation.  When the resolution is
decreased, the $f$-range where $\vdel(f)$ closely matches $\vtrue(f)$
narrows gradually.

As argued above, the systematic deviations at the low-$f$ end are
probably a result of the finite phase-space volume occupied by the
particles. As the number of particles is increased, a larger portion
of phase-space is covered, enabling the reconstruction of lower values
of $\fdel$.

The high-$f$ systematic deviations can be qualitatively understood
using the Voronoi model and its convolution formula (\ref{eq:av-vf}).
Since the DTFE uses a finite number of particles to recover the $f$
field, there must be an upper cutoff, $f_1$, for the spectrum of $f$
values that the DTFE can produce in principle.  As far as the DTFE is
concerned, the true $v(f)$ has an effective cutoff at $f_1$. If we
plug this ``true" truncated $v(f)$ into the convolution formula
(\ref{eq:av-vf}), assume a distribution function $p(\cdot)$ with a
width of about 1 decade, and set $f_1\simeq 50$, we recover the
qualitative behaviour of the measured $\vdel(f)$ for the $N=10^6$ mock
samples with $\sigv$ and the three values of $\alpha$.  This simple
picture is, however, an oversimplification because the high-$f$ region
is also effected by statistical fluctuations due to the low number of
particles there.

The convolution formula can also explain the \emph{overestimation} by
the $\vdel(f)$ in the high-$f$ regions of the mock systems with
$\sigc$.  In these systems, the transition between the low-$f$
power-law and the steep high-$f$ decline is rather sharp; it occurs
over a scale comparable to the width of the window function
$p(\cdot)$.  As a result, the $\vdel(f)$ measurement at a high-$f$
includes contributions from the higher $v(f)$ values at lower $f$.
This becomes very apparent in the case with $\alpha=0$ and $\sigc$, in
which $f$ has an upper bound of $f\simeq 63$. While the true $v(f)$
vanishes for all $f$ values larger than this limit, the DTFE-measured
$v(f)$ vanishes only at an $f$ value that is an order-of-magnitude
larger, due to the 1-decade width of the assumed probability
distribution function $p(\cdot)$.

\subsection{Statistical Errors in $\vdel(f)$.}
 
The Voronoi model can also be used to estimate the \emph{statistical}
errors in $\vvor(f)$. Strictly speaking, $\vvor(f)$ is an ill-defined
random variable, as it is measured by differentiating $\Vvor(f)$,
which is a super-position of Heaviside step functions, and as such
$\vvor(f)$ is a sum of Dirac delta functions. Much like a white-noise
process, its variance is infinite.  In practise, however, we always
compute $\vvor(f)$ by differentiating a \emph{smoothed} version of
$\Vvor(f)$ (using a spline, for example). Therefore, we may expect the
statistical error in $\vvor(f)$ to be comparable to the statistical
error in $\Vvor(f)$. The latter can be estimated in a way similar to
how we estimated the average of $\Vvor(f)$.

To calculate $\left<[\Delta \Vvor(f)]^2\right> =
\left<V^2_{vor}(f)\right> - \left<\Vvor(f)\right>^2$, we can use
the definition of $V_\alpha(f)$ to write
\begin{equation}
  \left<V^2_{vor}(f)\right> = \sum_{\alpha\ne\beta} 
    \left<V_\alpha(f)V_\beta(f)\right> 
   + \sum_\alpha \left<V^2_\alpha(f)\right> \ .
\end{equation}
Assuming that $V_\alpha(f)$ is independent of $V_\beta(f)$, the
cross terms would cancel out from $\left<[\Delta
  \Vvor(f)]^2\right>$, leaving us with the upper limit
\begin{equation}
  \left<[\Delta \Vvor(f)]^2\right> \le 
   \sum_\alpha \left<V^2_\alpha(f)\right> \ .
\end{equation}

Using arguments similar to those used for calculating
$\left<V_\alpha(f)\right>$, one can show that
\begin{eqnarray}
  \left<V^2_\alpha(f)\right> 
  &\!=\!&
      m \int_{\omega_\alpha} d\Bx d\Bv \int_{f}^\infty 
      p[\fvor/f(\Bx,\Bv)]  f^{-2}_{vor}\, d\fvor \\
  &\!\le\!& \frac{m}{f} \int_{\omega_\alpha} d\Bx d\Bv \int_{f}^\infty 
     p[\fvor/f(\Bx,\Bv)]  f^{-1}_{vor}\, d\fvor \\
  &\!=\!& \frac{m}{f}\left<V_\alpha(f)\right> \ .
\end{eqnarray}
Therefore, 
\begin{equation}
  \left<[\Delta \Vvor(f)]^2\right> \le 
      \frac{m}{f}\left<\Vvor(f)\right> \ ,
\end{equation}
and the relative error $\Delta(f)$ is given by
\begin{equation}
  \label{eq:stat-error}
  \Delta(f) \le \left(\frac{m}{f \left<\Vvor(f)\right>}
    \right)^{1/2} \ .
\end{equation}
Plugging $m=M/N$ into the formula above, where $M$ is the total mass
of the system and $N$ the total number of particles, we recover the
common large-numbers limit $\Delta(f) \propto 1/\sqrt{N}$.

\begin{figure}
\centerline{ \hbox{
\epsfig{file=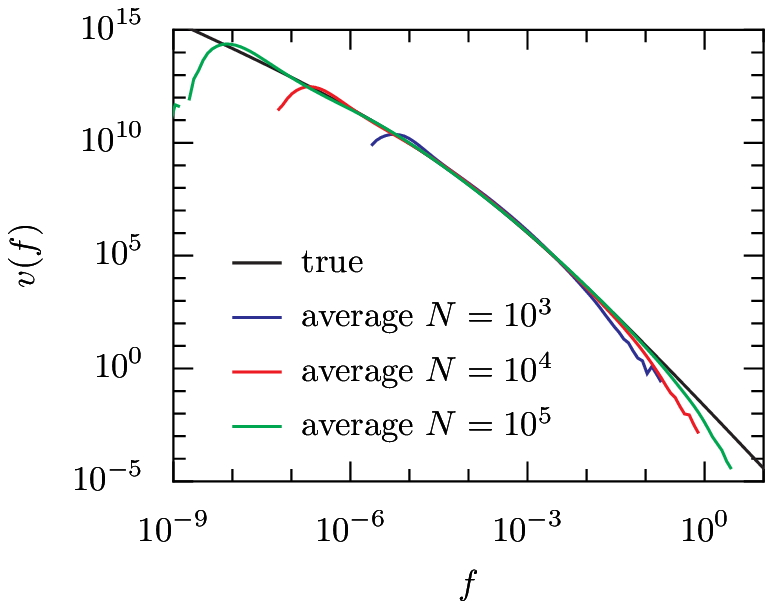,width=0.45\textwidth}}}
\centerline{ \hbox{
\epsfig{file=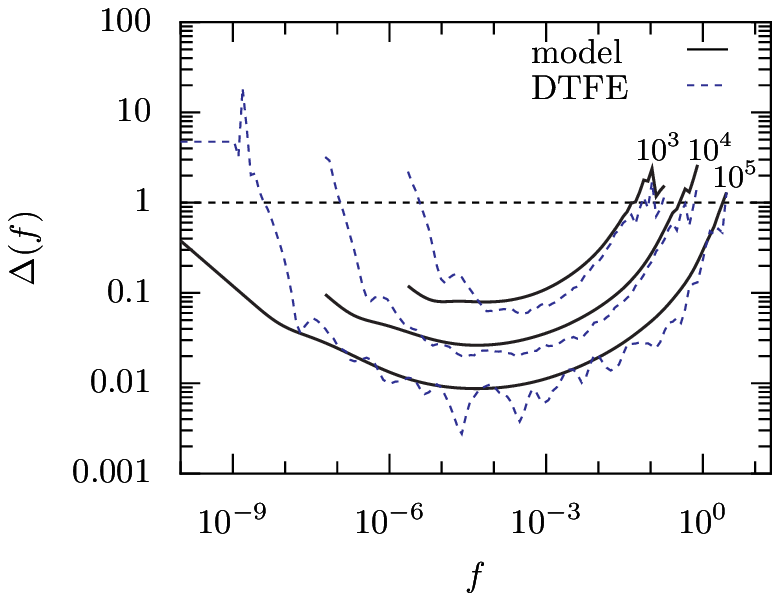,width=0.45\textwidth}}}
\caption{The average (top) and relative error (bottom)
  in $v(f)$ as measured in the mock system ($\alpha=1$, $\sigv$)
  sampled with three different resolutions.  The relative errors are
  compared to the analytic prediction of \Eq{eq:stat-error}.
} 
\label{fig:stat-errors}
\end{figure}

To check how good this estimate is for the DTFE-measured $v(f)$, we
have measured the $v(f)$ of 100 realisations of the $\alpha=1$ $\sigv$
mock system with $N=10^3$ and $N=10^4$ particles, and 30 realisations
with $N=10^5$ particles.  From these measurements we computed the true
relative error in $v(f)$ [with respect to the average of the
DTFE-measured $v(f)$, not with respect to the exact $v(f)$], and
compared it to the prediction of \Eq{eq:stat-error}.
Figure~\ref{fig:stat-errors} shows the comparison for the three
resolutions. We see that \Eq{eq:stat-error} performs well as an upper
bound for the statistical errors, except for the low-$f$ region.  In
that region the cumulative $V(f)$ approaches a constant as $f\to 0$,
due to the finite phase-space volume of the halo. This introduces
fluctuations to $v(f)$ as a result of the numerical differentiation in
\Eq{eq:vf-Vf}.

However, it is interesting to notice that whenever the statistical
errors in $\vdel(f)$ become important, they are overwhelmed by the
low-$f$ or high-$f$ systematic errors. In that respect, the
statistical errors in $\vdel(f)$ are of no big relevance.

\label{lastpage}
\end{document}